\documentclass[10pt,journal,compsoc]{IEEEtran}
\IEEEoverridecommandlockouts

\usepackage{amsmath, amssymb, mathrsfs, amsthm}
\usepackage[autostyle]{csquotes}
\usepackage{graphicx, comment, url}
\usepackage{listings,wrapfig,enumitem}
\usepackage{float, parskip}
\usepackage{color, xcolor, cases}
\usepackage{subfig, fancyhdr}
\usepackage[numbers,sort, compress]{natbib}
\usepackage[english]{babel}
\usepackage[nodisplayskipstretch]{setspace}
\usepackage{bbm} 

\usepackage{breakurl}
\usepackage[breaklinks, colorlinks,
    linkcolor={red!70!black},
    citecolor={blue!20!green},
    urlcolor={blue!80}]{hyperref}
\usepackage[hang,flushmargin]{footmisc}

\makeatletter
\def\footnoterule{\kern-3\p@
  \hrule \@width 1.5in \kern 2.6\p@} 
\makeatother

\newtheorem{thm}{\vspace{-6pt}\\ Theorem}
\newtheorem{defn}{\vspace{-6pt}\\ Definition}

\newtheorem{cor}[thm]{\vspace{-6pt}\\ Corollary}

\newtheorem{problem}{\vspace{-6pt}\\ Problem}

\newcommand{\s}{\mathbf{s}}
\newcommand{\M}{\mathcal{M}}
\newcommand{\p}{\mathbf{p}}
\newcommand{\n}{\mathbf{n}}
\newcommand{\X}{\mathbf{X}}
\newcommand{\x}{\mathbf{x}}
\newcommand{\Y}{\mathbf{Y}}
\newcommand{\Z}{\mathbf{Z}}

\newcommand{\C}{\mathcal{C}}
\newcommand{\bs}{\boldsymbol}
\allowdisplaybreaks
\def\BibTeX{{\rm B\kern-.05em{\sc i\kern-.025em b}\kern-.08em
    T\kern-.1667em\lower.7ex\hbox{E}\kern-.125emX}}

\begin{document}
\title{\huge Disclosure Risk from Homogeneity Attack in Differentially Privately Sanitized Frequency Distribution\\
\author{Fang Liu,  Xingyuan Zhao
\thanks{F. Liu is Professor (e-mail: Fang.Liu.131@nd.edu) and X. Zhao  (e-mail: xzhao8@nd.edu)  is a doctoral student in the Department of Applied and Computational Mathematics and Statistics, University of Notre Dame, Notre Dame, IN, 46556, USA.  This work was supported by NSF Award \#1717417.}\vspace{-24pt}
}}
\maketitle
\vspace{-54pt}
\begin{abstract}
Differential privacy (DP)  provides a robust model to achieve privacy guarantees for released information. We examine the protection potency of sanitized multi-dimensional frequency distributions (FDs) via DP mechanisms against homogeneity attack (HA). Adversaries can obtain the exact values on  sensitive attributes of their targets through HA without having to identify them from released data. We propose measures for disclosure risk (DR) from HA and derive closed-form relations between the privacy loss parameters and DR from HA. The availability of the closed-form relations will assist practitioners in understanding the abstract concepts of DP and privacy loss parameters by putting them in the context of a concrete privacy attack and offer a perspective for choosing privacy loss parameters when employing DP mechanisms. We apply the derived mathematical relations in real data to demonstrate the assessment of DR from HA on differentially privately sanitized FDs at various privacy loss parameters. The results suggest that relations between DR from HA and privacy loss are S-shaped; the former may not disappear even when privacy loss approaches 0. \vspace{-6pt}
\end{abstract} 
\begin{IEEEkeywords}
differential privacy;   disclosure risk;  homogeneity attack;  Gaussian mechanism;  Laplace mechanism;  privacy loss parameter; privacy budget; privacy-preserving
\end{IEEEkeywords}

\vspace{-9pt}
\section{Introduction}\vspace{-3pt}
\subsection{Background and Motivation}\vspace{-3pt}
Two common privacy risk types experienced by an individual during data sharing and information release are the re-identification risk and the  disclosure risk. Re-identification risk occurs when adversaries identify their targets in the released data and disclosure risk refers to the disclosure of private and sensitive information on individuals using the released information. Disclosure risk may occur to an individual without revealing the identity of that individual. The ultimate goal of re-identification, in many cases, is to learn sensitive information of targeted individuals. We focus on disclosure risk in this paper.

An indispensable step when developing a privacy protection and disclosure limitation technique is to  measure the effectiveness of the technique in protecting individual privacy or sensitive information. Post-hoc quantitative assessment of disclosure risk  is a long-standing research problem.  Various metrics assessing the risk have been developed; many rely on specific assumptions about the background knowledge or behaviors adversaries and the data  \citep{duncan1989risk, skinner2002measure, yancey2002disclosure, reiter2005estimating, hundepool2012statistical, hu2018bayesian}.

In recent years, differential privacy (DP) \citep{dwork2006calibrating} has become a mainstream notion in data privacy research and has been gaining popularity in industry, business, and government for data collection and release in practice.  Compared to the traditional posthoc disclosure risk approaches that assess the absolute risk, DP evaluates the incremental risk that the adversary learns additional information about his/her target from the released information on top of what he/she already knows, which can be controlled for a  pre-specified parameterized privacy loss. Different types of randomized mechanisms have been developed to release query results and statistics in general settings as well as for specific types of queries and analyses. Results released from a DP mechanism are immune to post-processing and future-proof; that is, there is no additional privacy leak about the individuals in the data where the results are based they are post-processed  (e.g. transformation) after release or if there is additional information on these individuals in the future from other sources. In addition, DP satisfies privacy loss composability and  amplification principles \citep{mcsherry2007mechanism, dwork2010boosting, kasiviswanathan2011can, abadi2016deep, mironov2017renyi}, making it attractive and convenient for  privacy loss tracking and accounting and privacy cost savings when multiple query results are released from the same data. 

Despite the desirable properties of DP and successful stories in its deployment in practice, the concept itself is rather abstract and appears less relatable and intuitive to practitioners compared to some of the privacy and disclosure risk measures associated with specific attack models that are easy to understand and interpret. In particular, there is no universal guideline on the choice of privacy loss parameters, which are key to implementing differentially private mechanisms in practice.  

The goal of this paper is to relate privacy loss parameters in DP and disclosure risk due to homogeneity attack (DR-HA). To our knowledge, no work exists on examining such relations.   HA is a common privacy attack model to disclose individual sensitive information without having to identify individuals in released data.  This type of attacks take advantage of the scenario where the values of a sensitive attribute are identical for a set of records that have the same identifiable information, often coarsened or anonymized. If an adversary's target belongs to that set, the value of his or her sensitive attribute can be attained, without correctly pointing out which record in that set is the target.

\vspace{-9pt}
\subsection{Related Work}\vspace{-3pt}
\citet{dwork2008differential} stated that ``the choice of $\epsilon$ is essentially a social question" and later interviewed   DP practitioners to understand the current DP practice in choosing privacy parameters and found no consensus \citet{dwork2019differential}.  Efforts have been made to connect the concept of DP and its privacy parameters with existing DR metrics and to examine the effectiveness of DP against various attacks (e.g., re-identification, disclosure, and reconstruction). For example, \citet{Clifton2011} considered the probability of re-identifying an individual from  a database and demonstrated the challenge of setting a proper value for privacy loss parameters. \citet{mcclure2012} used empirical studies to investigate probabilistic disclosure risk in differentially private univariate binary data. \citet{roth2014} proposed a  model to balance the interests of data analysts and data contributors and used the model to choose privacy loss parameters for some statistical analyses. \citet{abowd2015} addressed the question from an economic perspective 
and defined the optimal choice of $\epsilon$ through the formulation of the social planner’s problem.  \citet{nissim2017differential} provided an example on how privacy loss  may be interpreted as bounding the worst-case financial risk incurred by an individual participating in a research study. \citet{dwork2017exposed} examined the robustness of DP for releasing aggregate statistics protect privacy when facing reconstruction attacks and tracing attacks.  \citet{holohan2017k} designed the $(k,\epsilon)$-anonymity algorithm  for quasi-identifiers and evaluated its privacy protection against linking records in the perturbed data to the original records  using the nearest neighbor technique. \citet{chen2017evaluating} defined the risk of data disclosure based on noise estimation and determined the $\epsilon$ value for the Laplace mechanism using confidence for the noise estimation. \citet{chen2017data} proposed an algorithm for choosing privacy loss parameters, balancing disclosure risk and utility.

\vspace{-9pt}
\subsection{Our Contribution}\vspace{-3pt}
We examine the relationships between privacy loss parameters in DP and DR-HA on multi-dimensional Frequency Distributions (FDs), also known as data cubes and marginals, and  contingency tables and cross-tabulations in statistics. FDs are essentially count data and are of the commonly released data types by data collectors/curators. The availability of the closed-form relationships between DP privacy loss parameters and DR-HA will help practitioners better understand the concept of DP and the associated privacy loss parameters in the context of HA, and  provide a perspective for choosing privacy loss parameters when implementing differentially private mechanisms in data sanitization and releasing, along with other considerations. 
Our main contributions are summarized below. 
\vspace{-2pt}
\begin{itemize}[leftmargin=12pt, itemsep=-3pt]
\item We define several DR-HA measures for a multi-dimensional FD dataset. Users may report one or multiple DR-HA measures, depending on the assumptions they are willing to make (Section \ref{sec:DR-HA}).
\item We derive the mathematical relationships between DR-HA and privacy loss parameters for Laplace and Gaussian mechanisms in DP (Section \ref{sec:relation}).
\item We apply the closed-form mathematical relationships in real-life data and show how to leverage the relations to assist decision-making on privacy loss parameters when implementing DP mechanisms (Section \ref{sec:experiment}).
\item The theoretical and empirical results suggest that relations between DR-HA and privacy loss are S-shaped, the lower asymptote of which relates to the number of levels of a sensitive attribute; DR-HA may not disappear even when privacy loss approaches 0.
\end{itemize}

\vspace{-9pt}
\section{Preliminaries}\label{sec:prelim}\vspace{-3pt}
We overview some basic concepts of DP used in this work and introduce  HA in this section.

\vspace{-6pt}\subsection{Differential Privacy (DP)}\vspace{-7pt}
\begin{defn}[\textbf{$(\epsilon,\delta)$-DP} 
\citep{dwork2006calibrating,dwork2006our}]\label{defn:adp}
A randomized algorithm $\M$ is of $(\epsilon,\delta)$-DP if for  all neighboring dataset pairs $(D_1,D_2)$ differing by one record and for all subsets $\mathcal{S}\subseteq$ image$(\M)$,
\begin{equation}\label{eqn:DP}
\Pr(\mathcal{M}(D_1)\in \mathcal{S}) \leq e^{\epsilon} \Pr(\mathcal{M}(D_2)\in \mathcal{S})+\delta.
\end{equation}
\end{defn}
\vspace{-6pt}

DP is a mathematical concept that provides privacy guarantees for the individual in a dataset from which information is released. A small $\epsilon>0$ implies that the probability of identifying an individual or attaining sensitive information of a targeted individual based on the release data sanitized  by $\mathcal{M}$ is low. When $\delta=0$, $(\epsilon,\delta)$-DP reduces to pure $\epsilon$-DP. $\delta\in[0,1)$, usually a value close to 0 (inverse proportional to poly$(n)$) if not 0,  can be interpreted as the probability that the pure $\epsilon$-DP is violated. A similar concept to $(\epsilon, \delta)$-DP  is $(\epsilon, \delta)$-probabilistic DP (pDP), given below.
\begin{defn}[\textbf{$(\epsilon,\delta)$-probabilistic DP} 
\citep{machanavajjhala2008privacy}]\label{defn:pdp}
A randomized algorithm $\M$ satisfies $(\epsilon,\delta)$-probabilistic DP if
\begin{equation}\label{eqn:pDP}
\Pr\left(\bigg|\log\left(\frac{\Pr(\M(D_1))\in \mathcal{S})}{\Pr(\M(D_2))\in \mathcal{S})}\right)\bigg|> \epsilon\right)\leq \delta
\end{equation}
for all neighboring datasets pairs $(D_1,D_2)$ and  all $\mathcal{S}\subseteq image(\M)$.
\end{defn}
Various differentially private randomized mechanisms have been developed to sanitize information.  The  Laplace mechanism  and Gaussian mechanism are two popular choices for sanitizing numerical queries, both of which are based on the concept of global sensitivity.
\begin{defn}[$\ell_p$ global sensitivity (GS)] \citep{liu2018generalized}]
The $\ell_p$ GS of query $\s$ is\vspace{-3pt}
$$\Delta_p(\s)=\max_{D_1,D_2,|D_1\setminus D_2|=1} ||\s(D_1)-\s(D_2)||_p\mbox{ for } p>0.$$\vspace{-12pt}
\end{defn}
The $\ell_p$ GS measures the largest change in $\s$ between all neighboring dataset pairs ($|D_1\setminus D_2|=1$). The commonly used $\ell_p$ GS is the $\ell_1$ GS at $p=1$, on which the Laplace mechanism is based, and the $\ell_2$ GS at $p=2$, on which the Gaussian mechanism is based. 

Let $\s=\{s_j\}_{j=1,\ldots,r}$. The Laplace mechanism \citep{dwork2006calibrating} sanitizes $\s$ by adding  Laplace noise to it. That is, $\tilde{s}_j\!=\!s_j\!+\!e_j,\mbox{ where }e_j\!\sim\!\mbox{Lap}(0,\Delta_1(\s)/\epsilon)$
independently for $j\!=\!1,\ldots,r$. There are two types of the  Gaussian mechanism, satisfying $(\epsilon,\delta)$-DP and $(\epsilon,\delta)$-pDP, respectively. Similar to the Laplace mechanism, the sanitized statistic is $\tilde{s}_j\!=\!s_j\!+\!e_j$ for $j\!=\!1,\ldots,r$, but $e_j\!\sim\! \mathcal{N}(0,\sigma^2)$, where 
\begin{align}
\sigma \geq c\cdot\Delta_2(\s)/ \epsilon \mbox{ with $\epsilon<1$ and  $c^2>2\log(1.25/\delta)$} \label{eqn:gaussian}\\
\sigma\geq (2\epsilon)^{-1} \Delta_2(\s)\left(\sqrt{(\Phi^{-1}(\delta/2))^2+2\epsilon }-\Phi^{-1}(\delta/2)\right)\label{eqn:gaussianp}
\end{align}  \vspace{-12pt}

for $(\epsilon,\delta)$-DP \citep{dwork2014algorithmic} and $(\epsilon,\delta)$-pDP \citep{liu2018generalized}, respectively,
where $\Phi^{-1}$ is the inverse cumulative density function of the standard normal distribution.

\vspace{-12pt}
\subsection{Homogeneity Attack (HA)}\vspace{-3pt}
Before we introduce HA, we first present two definitions necessary for understanding HA. The first is \emph{quasi-identifiers}  (QIDs) \citep{dalenius1986finding}.  QIDs are not unique identifiers (e.g., social security numbers) but contain identifiable information that is sufficiently correlated with an individual and may lead to a unique identifier after being combined with other QIDs. Demographic attributes, such as age, race, gender, and geographical information, are  regarded as QIDs.  Adversaries often have exogenous knowledge of QIDs. The second is \emph{sensitive attributes}.  Sensitive attributes are attributes that contain sensitive information about individuals, such as income, medical history, criminal records, etc. These sensitive attributes are often of interest to adversaries who may launch different types of attacks on released information to disclose their values. We refer to this type of privacy risk as the disclosure risk (DR)  in this work. 

HA occurs where the values on a sensitive attribute are identical for a set of multiple records. Table  \ref{tab:HA} presents an example on a FD dataset subject to HA. The data contains 12  individual records  ``race'' and ``age'' can be regarded as QIDs while ``medical condition'' is a sensitive attribute the information about which is accurately presented in the data.  Records 1 to 4 share the same set of QIDs, so do records 5 to 8, and records 9 to 12, respectively. The group containing records 9 to 12 are also \emph{homogeneous} on ``medical condition''.  Suppose an adversary knows his/her target is in the dataset and wants to learn the medical condition of the target who is white and aged $\ge50$ years old. Despite the lack of knowledge on which record from 9 to 12 is his/her target, he/she still learns that the target has diabetes via HA.
\begin{table}[!h]\vspace{-3pt}
\centering
\caption{An example dataset subject to HA}\label{tab:HA}\vspace{-6pt}
\resizebox{0.8\columnwidth}{!}{
\begin{tabular}{cccc}
\hline	
& \multicolumn{2}{c}{quasi-identifier} & sensitive attribute\\
\cline{2-4}
ID & race & age (year) & medical condition\\
\hline
1 & black &	$<40$ & heart disease\\
2 & black & $<40$ & heart disease\\
3 & black  & $<40$ & cancer\\
4 & black  & $<40$ & cancer\\
\hline
5 & Hispanic & $[40,50)$ & cancer\\
6 &	Hispanic &  $[40,50)$ & diabetes\\
7 &	Hispanic &  $[40,50)$ & heart disease\\
8 &	Hispanic &  $[40,50)$ &  heart disease	\\
\hline
9 &	white & $\ge50$ &  diabetes\\
10 & white & $\ge50$ &  diabetes\\
11 & white & $\ge50$ &  diabetes\\	
12 & white & $\ge50$ &  diabetes\\	
\hline
\end{tabular}}
\vspace{-12pt} \end{table}

\vspace{-9pt}
\section{Disclosure Risk from Homogeneity Attack (DR-HA) on  Sanitized FDs}\label{sec:DR-HA}\vspace{-3pt}
\subsection{Problem Setting and Problem  Statement} \label{sec:problem} \vspace{-3pt}
We focus on  FDs (e.g., multi-dimensional histograms,  contingency tables). FDs are a common data type released by data curators (e.g., the US Census). We present the definitions of homogeneous and heterogeneous cells in FDs first and then state the problem we aim to solve.

\vspace{-3pt}\begin{defn}[\textbf{homogeneous cell}] \label{def:homo}
In a dataset of $n$ records, attributes $\X$ contain $p\ge1$ QIDs and $\Y$ comprise $q\ge1$ sensitive attributes. The cross-tabulation of $\X$ is indexed by $i$ with label $\x_i$.  A cell in the cross-tabulation of $\X$ is  a \emph{homogeneous cell} with respect to $Y_j$ for  $j\!=\!1,\ldots,q$, if it is non-empty and all records in the cell have the same value for $Y_j$; and it is denoted by  $\mathcal{H}(\x_i,y_{ij})$ with $\x_i$ and $y_{ij}$ referring to the labels of $\X$ and $Y_j$ of the cell, respectively. If the cell is homogeneous for every $Y_j$ for $j\!=\!1,\ldots,q$, it is a \emph{complete homogeneous cell}; otherwise, it is a {partial homogeneous cell}.
\end{defn}\vspace{-3pt}

\begin{defn}[\textbf{heterogeneous cell}] \label{def:hetero}
In a dataset of $n$ records, $\X$ contains $p\ge1$ QIDs and $\Y$ comprise $q\ge1$ sensitive attributes. A cell in the cross-tabulation of $\X$ is a \emph{heterogeneous cell}  with respect to $Y_j$  for  $j=1,\ldots,q$ if it is non-empty and there are at least two records in this cell having different values on $Y_j$. If the cell is heterogeneous for every $Y_j$ for $j=1,\ldots,q$, then it is a \emph{complete heterogeneous cell}. 
\end{defn}
For a given $Y_j$, a cell in the cross-tabulation formed by $\X$ is either homogeneous or heterogeneous if it is non-empty.  When $q\ge2$, a cell can be completely homogeneous, completely heterogeneous, or partially homogeneous.  Definition \ref{def:homo} covers the scenario of sample uniqueness, which describes the situation where a cell formed by cross-tabulation of $\X$ contains only a single record. Sample uniqueness is a special but trivial case of a homogeneous cell. 

Definitions \ref{def:homo} and \ref{def:hetero} apply to both original data and sanitized or anonymized data. The data in Table  \ref{tab:HA} contain three non-empty cells formed by QIDs ``race''  and ``age''. The 4 records in the cell (race = white; age $\ge50$) have the same value on ``condition'' and this cell is thus homogeneous with respect to ``condition'' per Definition \ref{def:homo}. The 4 records in the cell (zip code = Hispanic; age $\in[40,50)$) are different on ``medical condition'' and is thus a heterogeneous cell per Definition \ref{def:hetero}, so is the cell (race = black; age $<40$).  

All records in a homogeneous cell $\mathcal{H}$ are subject to HA that  may lead to the disclosure of sensitive information. If the information on $Y_j$ is accurately presented in the released data, then the adversary can learn information on $Y_j$ of an individual without actually identifying the individual.\footnote{If the sensitive value is not correctly presented in the released data because of data entry errors, measurement errors, missing values, intentional perturbation for privacy reasons, etc, then HA  does not necessarily lead to the disclosure of  sensitive information. But there might be other types of harm (e.g. social harm) if the adversary disseminates the wrong information, claiming the information is true, whether intentional or not. This type of harm can be mitigated if the data curator puts a disclaimer regarding the accuracy of individual-level information when publishing the data.} Original empty cells in the cross-tabulation of $\X$ do not pose DR as no individual is present in the sample data with such QID values. Zero counts in those cells may be sanitized if such QID+$Y$ combinations are deemed possible in the population even though they are not present in a particular sample dataset, but DR-HA remains null for these cells even if the sanitized cells may become non-empty and homogeneous in $Y$ as the ``imputed'' $Y$ values are  random.  
\begin{problem}\label{prob} 
Denote the non-empty cell set from the cross-tabulation of QIDs $\X$ in the original data by $\mathscr{C}_{\X}=\{\C_\X(\x_1),\ldots,\C_\X(\x_N)\}$ or $\mathscr{C}_{\X}\{\C_1,\ldots,\C_N\}$ for simplicity, where $\x_i$ represents the label of $\X$ in  cell $\C_i$ for $i=1,\ldots,N$; the cell sizes by $\n_{\X}=(n_1,\ldots, n_N)$; the sensitive attribute of the adversary's interest by $Y$ with $K$ distinct values ($1,\ldots,K$). Further cross-tabulation of $\mathscr{C}_{\X}$ and $Y$  generates $N\times K$ cells $\mathscr{C}=\{\mathcal{C}(\x_1,Y=1),\ldots,\mathcal{C}(\x_1,Y=K)\ldots,\mathcal{C}(\x_N,Y=1),\ldots,\mathcal{C}(\x_N,Y=K)\}$  with FD $\n=(\n_1,\ldots,\n_N)$, where  $\n_i=(n_{i1},\ldots,n_{iK})$ and $\sum_{k=1}^K n_{ik}=n_i$ for $i=1\ldots,N$. Suppose $\n$ is perturbed via a randomized mechanism $\M_{\bs\theta}$ with privacy loss parameters $\bs\theta$, leading to sanitized FD $\tilde{\n}$ over $\mathscr{C}$. What is the DR-HA on  $Y$ given $\tilde{\n}$?
\end{problem}
To address Problem \ref{prob}, we propose several measures for DR-HA in Section \ref{sec:DRHA} and derive their relations with privacy loss parameters in Section \ref{sec:relation}. 

\vspace{-9pt}\subsection{Measures of DR-HA}\label{sec:DRHA}\vspace{-5pt}
For a homogeneous cell  $\mathcal{H}(\x_i,Y=k|\n)$ in a dataset with $n$ records, further cross-tabulation of $\mathscr{C}_{\x}$ and $Y$ leads to FD $\n_i\!=\!\{n_{ik}\}_{k=1,\ldots,K}$ with  one non-zero  element and $(K-1)$ zero elements (i.e., $n_{ik}\!=\!n_i$ and $n_{ik'}\!=\!0$ for $k'\ne k$). After sanitization, there are four possible output scenarios $\tilde{\n}_i$ for $\n_i$, listed below. \vspace{-3pt}
\begin{itemize}[leftmargin=15pt]\setlength\itemsep{-3pt} 
\item Scenario 1: the sanitized cell remains  homogeneous with the same $Y$ value as the original $Y$ value $Y_k$; that is, $\mathcal{H}(\x_i,Y_k|\tilde{\n}_i)\ne\varnothing$ and $\tilde{\n}_i$ has one non-zero  element at the position $k$ and $K-1$ zero counts; i.e., $\tilde{n}_{ik}=\tilde{n}_{i}$ and $\tilde{n}_{ik'}=0$ for $k'\ne k$;
\item Scenario 2: the sanitized cell remains homogeneous but with a different  $Y$ value $Y_{\tilde{k}}$ other than the original $Y$ value $Y_k$; that is, $\mathcal{H}(\x_i,Y_{\tilde{k}}|\tilde{\n}_i)\ne\varnothing$, where $\tilde{k}\ne k$, and $\tilde{\n}_i$ has one non-zero  element at the position $\tilde{k}$ and $K-1$ zero elements; i.e., $\tilde{n}_{ik}=\tilde{n}_i$ and $\tilde{n}_{ik'}=0$ for $k'\ne \tilde{k}$. 
\item Scenario 3:$\!$ the sanitized cell becomes heterogeneous with at least two different $Y\!$ values. 
\item Scenario 4: $\tilde{\n}_i=\mathbf{0}$, i.e., $\C_{\X}(\x_i|\tilde{\n}_i)=\varnothing$.
\end{itemize} 
In summary, DR-HA still exists in Scenario 1, and disappears in Scenarios 2 to 4 for different reasons.

For a heterogeneous cell $\C_i$, its $\n_i$ has at least two  non-zero elements for different values of $Y$. After the sanitization , there are four possible outputs for of $\n_i$, listed below. \vspace{-3pt}
\begin{itemize}[leftmargin=12pt]\setlength\itemsep{-3pt} 
\item Scenario 5: $\C_{\X}(\x_i|\tilde{\n}_i)$ remains heterogeneous and $\tilde{\n}_i$ does not have to match $\n_i$ in either position or values.
\item Scenario 6: $\tilde{\n}_i=\mathbf{0}$  (i.e., $\C_{\X}(\x_i|\tilde{\n})=\varnothing$).
\item Scenario 7: the sanitized cell becomes homogeneous $\mathcal{H}(\x_i,Y_{\tilde{k}}|\tilde{\n}_i)$ and $\tilde{k}$ does not belong to set of $Y$ values in the original cell $\C_{\X}(\x_i|\n)$.
\item Scenario 8: the sanitized cell becomes homogeneous $\mathcal{H}(\x_i,Y_{\tilde{k}}|\tilde{\n}_i)$ and $\tilde{k}$ is one of the $Y$ values in the original cell $\C_{\X}(\x_i|\n)$.
\end{itemize}\vspace{-1pt}
In Scenarios 5 and 6,  there is obviously no DR-HA. In Scenarios 7 and 8, the cell becomes homogeneous after sanitization; whether this triggers DR-HA depends on whether any of the original $Y$ values in $\C_i$ remain after sanitization. In Scenario 7, the original $Y$ values in cell $\C_i$ are replaced by a new $Y$ value. Even though the sanitized cell is homogeneous, the information on $Y$ is wrong for the records in this cell and thus there is no DR-HA. In Scenario 8,  one of the original $Y$ values in the cell remains after sanitization and there is DR-HA  for the records in the cell whose original $Y$ value remains unchanged but not for those whose original $Y$ values disappear after sanitization since the released $Y$ information for the latter group is wrong. Though the net effect on DR in Scenario 8 is complicated and depends on the relative frequencies of these two groups, we take a conservative approach and treat this cell as being subject to DR-HA, along similar lines of ``worst case'' or ``upper bound''. To  better understand this, consider the following example. Suppose $Y$ is binary and cell $\C_i$ with QID $\x_i$ contains 100 records, 5 with label $Y\!=\!0$ and 95 with label $Y\!=\!1$ in the original FD. After sanitization, the cell becomes homogeneous with label $Y\!=\!1$. Though the sanitized information on $Y$ is not 100\% accurate in this cell, releasing the information leads to disclosure of the true $Y$ values for 95\% of the original records with QID $\x_i$ band provides the wrong $Y$ information for 5\% of the original records with QID $\x_i$. Our upper bound approach assumes all records in the cell are subject to HA though the $Y$  disclosed information is not accurate for a small portion of the records. The more unbalanced in terms of $Y$ values in a cell is, the more likely Scenario 8 will occur in that cell.

In summary, DR-HA exists in both the original and sanitized FDs in Scenario 1; exists in the original FD but disappears after sanitization in Scenarios 2 to 4; does not exist in either the original or sanitized FDs in Scenarios 5 to 7; and goes from zero to non-zero after sanitization in Scenario 8, partially due to the ``upper bound'' approach we adopt to be conservative and for technical simplicity.  

\begin{figure}[!htb]
\vspace{-12pt}\centering
\includegraphics[width=0.95\columnwidth]{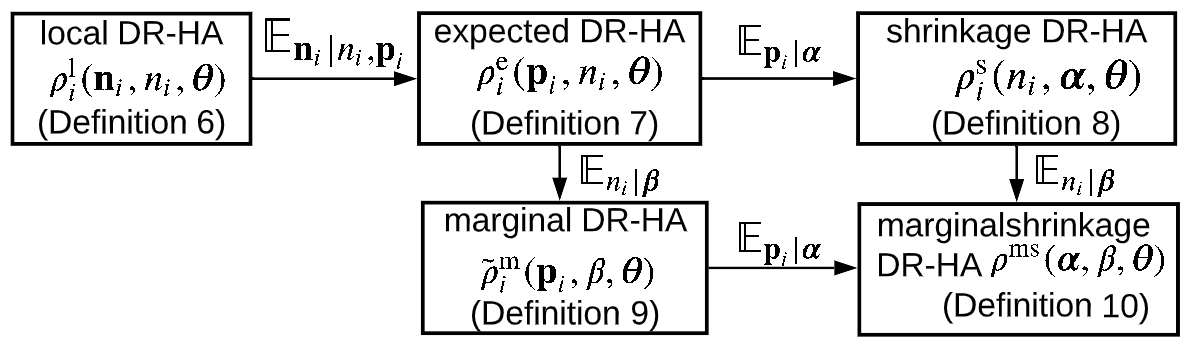} \vspace{-3pt}
\caption{Definitions of DR-HA.  \emph{local DR-HA} is the expectation of the DR-HA in a cell $\C_i$ over a randomized mechanism $\M_{\bs{\theta}}$ given FD $\n_i$;  \emph{expected DR-HA} is the expectation of local DR-HA over the distribution of $\n_i$ given cell size $n_i$;  \emph{shrinkage DR-HA}  is the expected value of expected DR-HA over a prior on $\p_i$; \emph{marginal DR-HA} is the expectation of expected DR-HA over the distribution of $n_i$; and \emph{shrinkage marginal DR-HA} is the expectation of shrinkage DR-HA over the distribution of $n_i$, or equivalently, the expectation of marginal DR-HA over a prior  on $\p_i$.}\label{fig:def} \vspace{-12pt}  
\end{figure}

The analysis of the eight scenarios above leads to five definitions on DR-HA in FDs after  sanitization (Definitions \ref{def:dr.c0} to \ref{def:dr.c4}). The definitions more or less form a sequence, as shown in Figure \ref{fig:def}.

\begin{defn}[\textbf{local DR-HA in a cell}] \label{def:dr.c0}
In the setting of Problem \ref{prob}, let $\mathcal{Y}_{i}$ denote the original set of the $Y$ labels in cell $\C_{\X}(\x_i)$ with non-zero frequencies. The local DR-HA $\rho^{\text{l}}_i\in[0,1]$ in $\C_{\X}(\x_i)$ after sanitization via randomized mechanism $\M_{\bs\theta}$ is \vspace{-4pt}
\begin{align}
&\rho^{\text{l}}_i\!=\!\rho_i(\n_i,\bs\theta)\label{eqn:untight0}\\
&\begin{cases}
=\Pr_{\M_{\bs\theta}}(\exists\; \tilde{k}\!\in\!\mathcal{Y}_i\!:\!\mathcal{H}(\x_i,Y_{\tilde{k}}|\tilde{\n}_i)\!\ne\! \varnothing) & \mbox{if }|\mathcal{Y}_i|\!=\!1\\
<\Pr_{\M_{\bs\theta}}(\exists\;\tilde{k}\!\in\!\mathcal{Y}_i\!:\!\mathcal{H}(\x_i,Y_{\tilde{k}}|\tilde{\n}_i) \!\ne\!\varnothing) & \mbox{if }|\mathcal{Y}_i|>1\!\label{eqn:untight}
\end{cases}
\end{align}
\end{defn}
\vspace{-6pt}
$|\mathcal{Y}_i|\!=\!1$  and $|\mathcal{Y}_i|\!>\!1$  refers to that $\C_{\X}(\x_i)$ is a homogeneous cell and  heterogeneous cell, respectively. The first equation  in Eq (\ref{eqn:untight}) corresponds to Scenario 1 and the second inequality yields the upper bound for Scenario 8 that is not tight (thus  the sign ``$<$").  Attaining a tight bound is possible if one defines what local DR-HA is when $|\mathcal{Y}_i|>1$. For example, we could let
\begin{equation}\label{eqn:tight}
\textstyle 
\rho^{\text{l}}_i\!\triangleq\!(n_{i\tilde{k}}/n_i)\cdot\Pr_{\M_{\bs\theta}}(\exists\; \tilde{k}\!\in\!\mathcal{Y}_i\!:\!\mathcal{H}(\x_i,Y_{\tilde{k}}|\tilde{\n}_i)\!\ne\!\varnothing);
\end{equation}
that is,  $(n_{i\tilde{k}}/n_i)\cdot 100\%$ of the original records in cell $\C_{\X}(\x_i)$ are subject to DR-HA, where $n_{i\tilde{k}}$ is the number of records whose $Y$ values do not change after the sanitization, to replace the upper bound as in  Eq (\ref{eqn:untight}). However,  Eq \eqref{eqn:tight} would impose difficulty in the analytical derivation of the  DR-HA metrics in Definitions \ref{def:dr.c1} to \ref{def:dr.c4}. In addition, Scenario 8 is only one out of 8 possible scenario in a FD and the probability it occurs is low unless $\n_i$ in a cell is highly unbalanced, $n_i$ is small, or the privacy loss is relatively low. Taken together, the upper bound in  Eq (\ref{eqn:untight}), though untight, is a good enough bound for assessing DR-HA for a whole dataset. 

\vspace{-3pt}\begin{defn}[\textbf{expected  DR-HA  in a cell}] \label{def:dr.c1}
In the same setting as Definition \ref{def:dr.c0}, assume $f(\n_i|n_i, \p_i)=$ multinomial$(n_i,\p_i)$, where $\p_i=(p_{i1},\ldots, p_{iK})$ and $\sum_k{p_{ik}}=1$. The expected DR-HA on $Y$ in cell $\C_{\X}(\x_i)$ after sanitization via mechanism $\M_{\bs\theta}$ is the expected value of the local DR-HA over $f(\n_i|n_i, \p_i)$, 
\vspace{-6pt}
\begin{align}\label{eqn:expected}
\rho^{\text{e}}_i&=\rho_i(n_i,\p_i,\bs\theta)=\mathbb{E}_{\n_i}(\rho^{\text{l}}_i)=
\int_{\n_i} \rho_i(\n_i,\bs\theta) f(\n_i|n_i,\p_i) d\n_i\notag\\
&<\Pr_{\n_i,\M_{\bs\theta}}(\exists \tilde{k}\in\mathcal{Y}_i: \mathcal{H}(\x_i,Y_{\tilde{k}}|\tilde{\n}_i) \ne \varnothing \cap |\mathcal{Y}_i|\!=\!1)+\notag\\
&\quad\Pr_{\n_i,\M_{\bs\theta}}(\exists \tilde{k}\in\mathcal{Y}_i:
\mathcal{H}(\x_i,Y_{\tilde{k}}|\tilde{\n}_I) \ne \varnothing \cap |\mathcal{Y}_i|\!>\!1).
\end{align}
\end{defn}\vspace{-6pt}
The first term in Eqn (\ref{eqn:expected}) covers Scenario 1 and the second term covers Scenario 8. $\rho^{\text{e}}_i$ is a function of $n_i,\p_i$, and $\bs\theta$ after taking expectation over over the distribution of $\n_i$. There are a couple of approaches to dealing with the unknown  $\p_i$ when estimating $\rho^{\text{e}}_i$. First, we can plug in the sample proportions $\hat{\p}_i=\n_i/n_i$. Second, we may assume a distribution on $\p_i$ and integrate it out, leading to shrinkage DR-HA in Definition \ref{def:dr.c2}.\vspace{-3pt}
\begin{defn}[\textbf{shrinkage DR-HA in a cell}] \label{def:dr.c2}
In the same setting as Definition \ref{def:dr.c1}, assume $\p_i\sim f(\p_i|\bs\alpha)$. The shrinkage DR-HA on $Y$ in cell $\C_{\X}(\x_i)$ after sanitization  via mechanism $\M_{\bs\theta}$ is the expected value of the expected DR-HA over $f(\p_i|\bs\alpha)$, \vspace{-3pt}
\begin{align}\label{eqn:shrinkage}
\!\!\!\rho^{\text{s}}_i\!=\!\rho_i(n_i, \bs\theta,\bs\alpha) \!=\!\mathbb{E}_{\p_i}(\rho^{\text{e}}_i)
\!\!=\!\!\int_{\p_i}\!\!\rho_i(n_i,\p_i,\bs\theta) f(\p_i|\bs\alpha) d\p_i.\!\!
\end{align}
\end{defn}
\vspace{-6pt} 
A natural choice for $f(\p_i|\bs\alpha)$ is Dirichlet$(\p_i|\bs\alpha)$ with hyper-parameter $\bs\alpha$.  One may specify $\bs\alpha$ based on prior knowledge, e.g., $\alpha_k\!=\!1$ for $k\!=\!1,\ldots,K$, or use empirical Bayes (EB) to determine $\bs\alpha$ given data $\n$. 

Definitions \ref{def:dr.c1} and \ref{def:dr.c2}  are conditional on $n_i$. In some cases, $n_i$  is fixed and pre-specified, such as in surveys or controlled experiments, where data are collected to achieve a pre-determined sample size per cell\footnote{For example, a $2\times2$ factorial design with factors gender (male or female) and ethnicity (Hispanic or not) is used to collect data on HIV status ($Y$). It aims to collect data $Y$ from 100 subjects in each of the 4 cells formed in the contingency table of gender and ethnicity. In this case, $n_i=100$ for $i=1,\ldots, 4$ is public knowledge.}. In other cases, $n_i$ is subject to sampling errors. We may further inter gates out the randomness around $n_i$ by assuming  a distribution $f(n_i|\bs\beta)$, leading to  marginal DR-HA in Definition \ref{def:dr.c3} and the marginal shrinkage DR-HA in Definition \ref{def:dr.c4}. 
\begin{defn}[\textbf{marginal DR-HA in a single cell}] \label{def:dr.c3}
In the same setting as Definition \ref{def:dr.c1}, assume $n_i\sim f(n_i|\bs\beta)$. The marginal DR-HA on $Y$ in cell $\C_{\X}(\x_i)$ after sanitization  via mechanism $\M_{\bs\theta}$ is the expected value of the expected DR-HA over $f(n_i|\bs\beta)$,\vspace{-4pt}
\begin{align}\label{eqn:marginal}
\!\!\rho_i^{\text{m}}\!=\!\rho(\p_i,\bs\beta,\bs\theta)
\!=\!\mathbb{E}_{n_i}(\rho^{\text{e}}_i)
\!=\!\!\int_{n_i}\!\!\rho_i(n_i,\p_i,\bs\theta) f(n_i|\bs\beta)dn_i.\!
\end{align}
\end{defn}
\vspace{-4pt} \begin{defn}[\textbf{marginal shrinkage DR-HA in a single cell}] \label{def:dr.c4}
In the same setting as Definition \ref{def:dr.c2}, assume $n_i\sim f(n_i|\bs\beta)$. The marginal shrinkage DR-HA on $Y$ in a cell after sanitization via $\M_{\bs\theta}$ is the expectation of the shrinkage DR-HA over of $f(n_i|\bs\beta)$,\vspace{-4pt}
\begin{align}\label{eqn:marginalshrinkage}
&\rho^{\text{ms}}\!=\!\rho(\bs\alpha,\bs\beta,\bs\theta)=\mathbb{E}_{n_i}(\rho^{\text{s}}_i)
\!=\!\mathbb{E}_{n_i,\p_i}(\rho^{\text{e}}_i)\notag\\
\!=\!&\int_{n_i}\!\int_{\p_i}\!\rho_i(n_i,\p_i,\bs\theta)f(\p_i|\bs\alpha) f(n_i|\bs\beta)d\p_idn_i.
\end{align}
\end{defn}\vspace{-5pt} 
$\rho^{\text{ms}}$ can be equivalently  defined as the expected value of $\rho^{\text{m}}$  over the distribution of $\p_i$. The unknown parameter $\bs\beta$ in Eqns \eqref{eqn:marginal} and \eqref{eqn:marginalshrinkage} can be estimated using  any appropriate inferential  approach given the observed data, such as method of moments (MoM), maximum likelihood estimation (MLE), or Bayesian inference. Note that parameters $\bs\beta$ and $\bs\alpha$  in $\rho_i^{\text{s}}, \rho_i^{\text{m}},$ and $\rho^{\text{ms}}$  are different in nature. $\bs\alpha$ is a hyperparameter that governs the distribution of the unknown parameter $\mathbf{p}_i$, whereas $\bs\beta$ is the parameter in the distribution of the observed $n_i$.  
Though $\rho^{\text{ms}}$ is defined for a single cell,  it  measures the DR-HA for a generic cell in any FD  sanitized by $\M_{\bs\theta}$ as long as the joint distribution of $\n_i, \p_i$, and $n_i$ is well approximated by multinomial$(\n_i|n_i,\p_i)f(\p_i|\bs\alpha)f(n_i|\bs\beta)$.  By contrast, $\rho_i^{\text{l}},\rho_i^{\text{e}}, \rho_i^{\text{m}}$, and $\rho_i^{\text{s}}$ are defined for cell $\C_i$. To assess DR-HA for a FD dataset that comprises multiple cells, we may average cell-level DR-HA across the cells in the data.\vspace{-3pt}
\begin{defn}[\textbf{average DR-HA}] \label{def:dr}
In the same setting as Definitions \ref{def:dr.c0} to \ref{def:dr.c3}, the unweighted and weighted empirical average DR-HA $\bar{\rho}_{\text{uw}}^*$ and $\bar{\rho}_{\text{w}}^*$ on $Y$ across $N$ cells formed by $\X\!$ after sanitization via randomized mechanism $\M$ are \vspace{-3pt}
\begin{align}
\begin{cases}\label{eqn:average}
\bar{\rho}_{\text{uw}}^*=\textstyle 
\!\sum_{i=1}^Nw_i\rho^*_i, & \mbox{ where }w_i=N^{-1}, \\
\bar{\rho}_{\text{w}}^*=\textstyle\sum_{i=1}^N w_i \rho^*_i,
&\mbox{ where }w_i= n_i/\sum_{i=1}^Nn_i, 
\end{cases}
\end{align} 
respectively; the superscript * can be  l, e, s, or m.
\end{defn}
The unweighted $\bar{\rho}_{\text{uw}}^*$ weighs every cell equally when it comes to aggregating DR-HA across all cells in an FD dataset whereas larger cells carry more weight (proportional to cell size) than smaller cells in the weighted $\bar{\rho}_{\text{w}}^*$.

\vspace{-12pt}\subsection{Choosing a DR-HA measure}\label{sec:choice}\vspace{-3pt}
Local DR-HA is ``local'' because it  measures the expected DR-HA in a cell sanitized  via $\M_{\bs\theta}$ given ``local'' data. 

Expected DR-HA is more ``global'' than local HR-HA as it integrates out the sampling error around $\n_i$ in a local cell   and measures DR-HA in a cell whose FD of $Y$ given a fixed cell size $n_i$ follows the same the distribution as $\n_i$.  

Marginal DR-HA further integrates out the sampling error around $n_i$ in the local data, and measures DR-HA in a cell whose FD of $Y$  given $n_i$ and the distribution of the cell size follows the same  distributions  as $\n_i$ and  $n_i$, respectively.  

Shrinkage DR-HA borrows information across cells to estimate $\p_i$  through a prior distribution $f(\p_i,|\bs{\alpha})$, known as ``shrinkage'' in Bayesian statistics, instead of estimating $\p_i$ separately in each cell as in expected DR-HA. Expected DR-HA can be treated as a special case of shrinkage DR-HA when $\bs{\alpha}$, the parameter that controls the shrinkage across cells, is set at a value that leads to no shrinkage. 

Lastly, shrinkage marginal DR-HA is the most ``global'' of all and integrates out all possible sources of sampling errors in the data.  

We provide the five DR-HA definitions for completeness. For practical implementation, if data curators are only interested in DR-HA in the local data per se, local DR-HA can be used without making distributional assumptions on the observed data; local DR-HA is also the easiest to calculate. If data curators aim to learn what DR-HA would be in observed-data-like datasets (following the same underlying distribution as the observed data), the other 4 measures can be considered, among which expected DR-HA is the least ``global'', but also the most straightforward to calculate and requires the least distributional assumptions.

\vspace{-6pt}\section{Relationship between DR-HA in sanitized FD and DP privacy loss parameters}\label{sec:relation}\vspace{-3pt}
The DR-HA definitions in Section \ref{sec:DRHA}  on FDs sanitized via a DP mechanism $\M_{\bs\theta}$ are generic and apply to any randomized $\M_{\bs\theta}$ with well-defined  $\bs\theta$. In this section, we derive closed-form relations between the DR-HA measures and the privacy loss parameters from the Laplace mechanism of $\epsilon$-DP (i.e. $\bs\theta\!=\!\epsilon$) and the Gaussian mechanisms of $(\epsilon,\delta)$-DP and $(\epsilon,\delta)$-pDP  (i.e. $\bs\theta\!=\!(\epsilon,\delta)$) to answer  Problem \ref{prob}.  The Laplace and Gaussian mechanisms are the most common and popular mechanisms for achieving DP in numerical query release, counts included. Other mechanisms exist for sanitizing counts with better utility (e.g., the geometric mechanism \cite{ghosh2009universally}). Given our goal is to quantify relationships between privacy loss parameters and DR-HA rather than focusing on utility, we choose to  study the popular Laplace and Gaussian mechanisms first and will explore relationships for other mechanisms in the future (see Section \ref{sec:discussion} for more discussion).

There are at least two benefits of having the relations in closed form. First, practitioners can apply the relations  to calculate DR-HA directly given $\bs\theta$, saving time and computational cost  on empirical evaluation of DR-HA otherwise; second, it can assist practitioners with interpreting  $\bs\theta$, choosing $\bs\theta$ in practical implementations, and evaluating the effectiveness of $\M_{\bs\theta}$ in the context of HA.  

The listing of the results is given in  Table \ref{tab:summary} and the detailed results are presented in Sec \ref{sec:laplace} and \ref{sec:gaussain}.
\begin{table}[!htb]
\centering\vspace{-3pt}
\caption{Results on the relationship between DR-HA and privacy loss parameters in DP mechanisms}\label{tab:summary}\vspace{-6pt}
\resizebox{1\columnwidth}{!}{
\begin{tabular}{l|l|l}
\hline
result & DR-HA ($\bar{\rho}^{\text{l}},\bar{\rho}^{\text{e}},\bar{\rho}^{\text{s}},\bar{\rho}^{\text{m}}, \rho^{\text{ms}}$) & DP mechanism \\
\hline
Theorem \ref{thm:rho.lap.K} & general & Laplace \\
Corollary \ref{cor:rho.lap.homo}  & all original cells are homogeneous& $\epsilon$-DP\\
%Corollary \ref{cor:rho.lap}$^\dagger$ & $K=2$  &\\
\hline
Theorem \ref{thm:rho.gau.K} & general & Gaussian \\
Corollary \ref{cor:rho.Gau.homo} &all original cells are homogeneous &  $(\epsilon,\delta)$-pDP,$(\epsilon,\delta)$-DP \\
%Corollary \ref{cor:rho.gau}$^\dagger$ & $K=2$ & $(\epsilon,\delta)$-DP\\
\hline
\end{tabular}}
\resizebox{1\columnwidth}{!}{
\begin{tabular}{l}
$^\dagger$ presented in the supplementary materials.$\mbox{\hspace{2.1in}}$\\
\hline
\end{tabular}}\vspace{-6pt}
\end{table}

\vspace{-6pt}\subsection{DR-HA in FD sanitized via Laplace mechanism}
\label{sec:laplace}\vspace{-9pt}
\begin{thm}[\textbf{relationship between DR-HA and $\epsilon$ in Laplace mechanism}]\label{thm:rho.lap.K}
Sensitive attribute $Y\!$ has $K\!\ge\!2$ distinct values. $n_{ik}$ is the frequency of its $k$-th value in cell $\C_i$ for $i\!=\!1,\ldots,N$ and $k\!=\!1,\ldots,K$; $\sum_{k=1}^{K}n_{ik}\!=\!n_i$. The Laplace mechanism of $\epsilon$-DP releases sanitized $\tilde{n}_{ik}\!=\!n_{ik}\!+\!e_{ik}$, where $e_{ik}\!\sim\!\mbox{Lap}(0,\epsilon^{-1})$. The average local DR-HA in Eqns \eqref{eqn:average} of the sanitized FD is 
\begin{align}
\!\!\!\!\bar{\rho}^{\text{l}}\!<\!
\mbox{\small $\big(1\!-\!\frac{1}{2}e^{-0.5\epsilon}$}&
\mbox{\small $\big)^{K\!-\!1}\!\sum_{i=1}^N\!\big\{\!w_i \mathbbm{1}(|\mathcal{Y}_i|\!=\!1)\big(1\!-\!\frac{1}{2}e^{(0.5-n_i)\epsilon}\big)$}\!\!\!\!\label{eqn:averagel.lap.K}\\
\!\!\!\!
\mbox{\small $+ w_i\mathbbm{1(|\mathcal{Y}_i|\!>\!1)}$}&
\mbox{\small $\frac{1}{2}(1\!-\!\frac{1}{2}e^{-0.5\epsilon})^{K\!-\!2}\big(e^{-0.5\epsilon}\!+\!e^{(1.5-n_i)\epsilon}\!-\!e^{(1-n_i)\epsilon}\!\big)\!\big\}$}.\notag
\end{align}
\normalsize
Assume $n_{ik}\sim$ multinomial$(n_i,\p_i)$, where  $\p_i\!=\!(p_{i,1},\ldots,p_{ik})$ represents the population proportions with $\sum_{k=1}^{K}p_{ik}=1$. Denote the sample estimate of $p_{ik}$ by $\hat{p}_{ik}=n_{ik}/n_i$. The plug-in estimate of the average expected DR-HA in Eqn \eqref{eqn:average} is \vspace{-3pt}
\begin{align}
\hat{\bar{\rho}}^{\text{e}}\!<& \textstyle \big(1\!-\!\frac{1}{2}e^{-0.5\epsilon}\big)^{K\!-\!1}\!\sum_{i=1}^N\!\big\{w_i\big(1-\frac{1}{2}e^{(0.5-n_i)\epsilon}\big)\sum_{k=1}^{K}\hat{p}_{ik}^{n_i}\big\}\notag\\
&\textstyle + (1\!-\!\frac{1}{2}e^{-0.5\epsilon})^{K-2}\!\sum_{i=1}^N\!\big\{w_i\mathbbm{1}(n_i\!\ge\!2)\times\label{eqn:averagee.lap.K}\\
&\textstyle\quad\frac{1}{2}\big(1\!-\!\sum_{k=1}^{K}\hat{p}_{ik}^{n_i}\big)\big(e^{-0.5\epsilon}+e^{(1.5-n_i)\epsilon}-e^{(1-n_i)\epsilon}\big)\big\}.\!\notag
\end{align}
Assume $\p_i\!\sim\!\mbox{Dirichlet}(\alpha_1,\ldots,\alpha_K)$; let $$\mbox{$A_i=\sum_{k=1}^{K}\frac{\Gamma(\sum_k\alpha_k)\Gamma(\alpha_{k}+n_i)}{\Gamma(\sum_k\alpha_k+n_i)\Gamma(\alpha_{k})}$},$$ 
the average shrinkage DR-HA  in Eqn \eqref{eqn:average} is
\begin{align}
&\textstyle \bar{\rho}^{\text{s}}\!<\!\left(1\!-\!\frac{1}{2}e^{-0.5\epsilon}\right)^{K-1}\sum_{i=1}^{N}\!\big\{w_iA_i\big(1\!-\!\frac{1}{2}e^{(0.5-n_i)\epsilon}\big)\big\}\notag\\
&\qquad\textstyle + (1\!-\!\frac{1}{2}e^{-0.5\epsilon})^{K-2}\!\sum_{i=1}^{N}\!\big\{w_i\mathbbm{1}(n_i\!\ge\!2)(1-A_i)\times\label{eqn:averages.lap.K} \\
&\qquad\quad\textstyle\frac{1}{2}(e^{-0.5\epsilon}\!+\!e^{(1.5-n_i)\epsilon}\!-\!e^{(1-n_i)\epsilon})\!\big\}.  \notag
\end{align}
Assume $n_i\sim f(n_i;\bs{\beta})$, the plugged-in estimate of the average marginal DR-HA  in Eqn  \eqref{eqn:average} is
\begin{align}
&\textstyle\bar{\hat{\rho}}^{\text{m}}\!<\!\big(1\!-\!\frac{1}{2}e^{-0.5\epsilon}\big)^{K\!-\!1}\!\sum_{i=1}^N\!\!\big\{w_if(n_i;\bs\beta)\!\big(1\!-\!\frac{1}{2}e^{(0.5-n_i)\epsilon}\big)\!\sum_{k=1}^{K}\hat{p}_{ik}^{n_i}\!\big\}\notag\\
&\!+\!\textstyle(1\!-\!\frac{1}{2}e^{-0.5\epsilon})^{K-2}\!\sum_{i=1}^N\!\big\{w_if(n_i;\bs\beta) \mathbbm{1}(n_i\!\ge\!2)\big(1\!-\!\sum_{k=1}^{K}\hat{p}_{ik}^{n_i}\!\big)
\notag\\
&\qquad\qquad\qquad\textstyle\times\frac{1}{2}\big(e^{-0.5\epsilon}+e^{(1.5-n_i)\epsilon}-e^{(1-n_i)\epsilon}\big)\big\}.\label{eqn:averagem.lap.K} 
\end{align}
\normalsize The marginal shrinkage DR-HA in Eqn \eqref{eqn:marginal} is
\begin{align}
&\rho^{\text{ms}}\!<\!\textstyle\big(1-\frac{1}{2}e^{-0.5\epsilon}\big)^{K-1}\!\sum_{n_i=1}^{\infty}\!\big\{f(n_i;\bs\beta)A_i(1\!-\!\frac{1}{2}e^{(0.5-n_i)\epsilon})\big\}\notag\\
&+\textstyle(1\!-\!\frac{1}{2}e^{-0.5\epsilon})^{K-2} \times\label{eqn:averagems.lap.K}\\
&\textstyle\sum_{n_i=2}^{\infty}\!\big\{f(n_i;\bs\beta)(1-A_i)0.5(\!e^{-0.5\epsilon}\!+\!e^{(1.5-n_i)\epsilon}\!-\!e^{(1-n_i)\epsilon})\!\big\}.\notag
\end{align}
\end{thm}
The proof is provided in the supplementary materials.  We note that the two summation terms in Eqns (\ref{eqn:averagel.lap.K}) to (\ref{eqn:averagems.lap.K}) correspond to Scenarios 1 and 8 in Sec \ref{sec:DRHA}, respectively. Both $(1\!-\!\frac{1}{2}e^{-0.5\epsilon})^{K-1}$ and $1\!-\!\frac{1}{2}e^{(0.5-n_i)\epsilon}$ increase in $\epsilon$ in the first term (Scenario 1). The  relationship of the second term  (Scenario 8) with $\epsilon$ is more complicated as $(1\!-\!\frac{1}{2}e^{-0.5\epsilon})^{K-2}$ increases and $(e^{-0.5\epsilon}+e^{(1.5-n_i)\epsilon}-e^{(1-n_i)\epsilon})$ decreases in $\epsilon$.   As $\epsilon\rightarrow\infty$, the cells return to their original homogeneous and heterogeneous forms, the first term becomes 1 and  the second term goes to 0, respectively.  

In terms of the specification of $f(n_i;\bs\beta)$, Poisson distribution $n_i\sim \mbox{Poisson}(\lambda)$ or negative binomial distribution  $n_i\sim \mbox{NegBin}(r,\lambda)$ are natural choices given that $n_i$ is count data, in which case $f(n_i;\bs\beta)$ in  Eqns (\ref{eqn:averagem.lap.K}) and (\ref{eqn:averagems.lap.K})  would be replaced by $\frac{e^{-\lambda}\lambda^{n_i}}{n_i!}$  and  $\binom{n_i+r-1}{r-1}(1-\lambda)^{n_i}\lambda^r$, respectively. $\bs\beta$ can be estimated via MoM, MLE, or Bayesian approaches based on the distribution assumption $f(n_i|\bs\beta)$, and then plugged in  Eqns (\ref{eqn:averagem.lap.K}) and  (\ref{eqn:averagems.lap.K}) to obtain $\bar{\hat{\rho}}^{\text{m}}$ and $\rho^{\text{ms}}$.  $\rho^{\text{ms}}$ in Eqn (\ref{eqn:averagems.lap.K}) involves summation over infinite terms. In practice, $n_i$ is always bounded and the truncated versions of $f(n_i|\beta, n_i\le n)$ can be used.  The hyperparameter $\bs\alpha$ in Dirichlet$(\alpha_1,\ldots,\alpha_K)$ can be specified or estimated using the EB approach (see the supplementary materials) and then plugged in Eqns (\ref{eqn:averages.lap.K}) and  (\ref{eqn:averagems.lap.K}) to estimate $\bar\rho^{\text{s}}$ and  $\rho^{\text{ms}}$. 

We examine two special cases of Theorem \ref{thm:rho.lap.K} when $K=2$ and when all the original cells are homogeneous, respectively. The results at  $K=2$ are presented  in the supplementary materials due to space limitation. In summary, the only component that involves $\epsilon$ in the second summation term in Eqns \eqref{eqn:averagee.lap.K} to \eqref{eqn:averagems.lap.K} is $(e^{-0.5\epsilon}+e^{(1.5-n_i)\epsilon}-e^{(1-n_i)\epsilon})$, which monotonically decreases in $\epsilon$ for any $n_i\geq2$. In other words, the upper bound for DR-HA increases for cells in Scenario 8  as $\epsilon$ decreases.    The results when all the original cells are homogeneous are presented in Corollary \ref{cor:rho.lap.homo}. Since Scenario 8 does not exist in this case (the reason behind the usage of an upper bound), we can obtain the exact DR-HA rather than using an upper bound. 
\vspace{-3pt}\begin{cor}\label{cor:rho.lap.homo}
If all the original cells are homogeneous, the average DR-HA measures in an FD sanitized via the Laplace mechanism of $\epsilon$-DP are
\begin{align}
\!\!\!\bar{\rho}^{\text{l}}\!=&\bar{\hat{\rho}}^{\text{e}}\!=\mbox{\small$\textstyle\big(1\!-\!\frac{1}{2}e^{-0.5\epsilon}\big)^{K\!-\!1}\!\sum_{i=1}^N\!\!\big\{\!w_i\big(1\!-\!\frac{1}{2}e^{(0.5-n_i)\epsilon}\big)\!\big\}$}\!\label{eqn:averagel.lap.homo.K}\\
\!\!\bar\rho^{\text{s}}=&\textstyle\big(1\!-\!\frac{1}{2}e^{-0.5\epsilon}\big)^{K-1}\!\sum_{i=1}^{N}\!\big\{w_iA_i\big(1\!-\!\frac{1}{2}e^{(0.5-n_i)\epsilon}\big)\!\big\}\label{eqn:averages.lap.homo.K}\\
\!\!\!\bar{\hat{\rho}}^{\text{m}}\!=&\mbox{\small$\textstyle(1\!-\!\frac{1}{2}e^{-0.5\epsilon})^{K-1} \sum_{n_i=1}^N\!\big\{w_if(n_i;\bs\beta)\big(1\!-\!\frac{1}{2}e^{(0.5-n_i)\epsilon}\big)\!\big\}$}\!\!\label{eqn:averagem.lap.homo.K}\\
\!\!\!\rho^{\text{ms}}\!=&\mbox{\small$\textstyle\big(\!1\!-\!\frac{1}{2}e^{-0.5\epsilon}\big)^{K-1}\sum_{n_i=1}^\infty\!\big\{A_if(n_i;\bs\beta)\big(\!1\!-\!\frac{1}{2}e^{(0.5-n_i)\epsilon}\!\big)\!\big\},$}\!\label{eqn:averagems.lap.homo.K}
\end{align}
respectively. All four measures $\bar{\hat{\rho}}^{\text{e}},\bar{\rho}^{\text{s}}, \bar{\hat{\rho}}^{\text{m}},\rho^{\text{ms}}\in(2^{-K},1)$.
\end{cor}
Eqns \eqref{eqn:averagel.lap.homo.K} to \eqref{eqn:averagems.lap.homo.K} can be obtained directly from Eqns \eqref{eqn:averagel.lap.K} and \eqref{eqn:averagems.lap.K} by dropping the second term in the summation, and further plugging in  $\bs{\hat{p}}_i\!=\!\{\hat{p}_{i,1}, \cdots, \hat{p}_{ik}\}=\{1, 0, \cdots, 0\}$ WLOG in Eqns \eqref{eqn:averagee.lap.K}  and \eqref{eqn:averagem.lap.K} (i.e., $K-1$ sample proportions are 0 and one is 1). In all cases,  the $<$ signs in Theorem \ref{thm:rho.lap.K} become $=$ due to the non-existence of scenario 8. There are several  take-away messages from Corollary \ref{cor:rho.lap.homo}. 
\begin{itemize}[leftmargin=12pt]\setlength\itemsep{-3pt} 
\item DR-HA is lower-bounded by $2^{-K}$ when all cells are homogeneous, regardless of $\epsilon,\n,$ and $N$ (e.g, when $K=2$, the lower bound is 25\%). In other words, no matter how small $\epsilon$ is, the Laplace mechanism can only lower the DR-HA  to $2^{-K}$ rather than 0 if all the original cells are homogeneous.  As $K$ increases, the lower bound approaches $0$ as it is more likely for a homogeneous cell  to become heterogeneous (i.e., the possibility of having at least two different labels of $Y$ in a cell increases) after sanitization.   
\item The larger $\epsilon$ is, the closer DR-HA is to 1 for a given $K$.
\item Given $\epsilon$, the smaller $n_i$ is, the lower DR-HA is as the sanitization has more impact on small cells than on large cells.  On the other hand,  $1-\frac{1}{2}e^{(0.5-n_i)\epsilon}$ is close to 1 even for not-so-large $n_i$ (e.g., it is $\ge0.985$ when $n_i\ge4$, $\ge0.959$ when $n_i\ge3$, $\ge0.888$ when $n_i\ge2$). This implies DR-HA is largely determined by $1-\frac{1}{2}e^{-0.5\epsilon}$ which is independent of the actual data information if all the original cells are homogeneous. This is also demonstrated in the experiments in Sec \ref{sec:experiment}, where the DR-HA is similar between two datasets where all the cells in the cross-tabulation of QIDs are homogeneous though they differ significantly in $n, N, n_i$.
\end{itemize}

\vspace{-9pt}
\subsection{DR-HA in FD sanitized via Gaussian mechanisms of \texorpdfstring{$(\epsilon,\delta)$}{}-DP and \texorpdfstring{$(\epsilon,\delta)$}{}-pDP}\label{sec:gaussain} \vspace{-9pt}
\begin{thm}[\textbf{relationship between DR-HA and privacy loss parameters $(\epsilon,\delta)$ in Gaussian mechanism}]\label{thm:rho.gau.K}
In the same setting as in Theorem \ref{thm:rho.lap.K}, the FD is sanitized via a Gaussian mechanism $\tilde{n}_{ik}\!=\!n_{ik}\!+\!e_{ik}$ with $e_{ik}\!\sim\!\mbox{N}(0,\sigma^2)$, where
$\sigma=\epsilon^{-1}\sqrt{2\ln{(1.25/\delta)}}\mbox{ with $\epsilon<1$}$
for the Gaussian mechanism of $(\epsilon,\delta)$-DP and 
$\sigma=(2\epsilon)^{-1}(\sqrt{(\Phi^{-1}(\delta/2))^2+2\epsilon }-\Phi^{-1}(\delta/2))$
for the Gaussian mechanism of $(\epsilon,\delta)$-pDP. Let $E_i=$\vspace{-6pt}
$$\textstyle \!\!\left(\!1\!-\!\mbox{erf}\big(\!\frac{1.5\!-\!n_i}{\sqrt{2}\sigma}\!\big)\!\right)\!\!\left(\!1\!+\!
\mbox{erf}\big(\frac{-0.5}{\sqrt{2}\sigma}\!\big)\!\right)\!+\!
\left(\!1\!+\!\mbox{erf}\big(\frac{1.5\!-\!n_i}{\sqrt{2}\sigma}\big)\!\right)\!\left(\!1\!+\!\mbox{erf}\big(\frac{0.5}{\sqrt{2}\sigma}\big)\!\right)\!,$$
\normalsize where erf() is the error function. The average local DR-HA  in Eqn \eqref{eqn:average} of the sanitized FD is \vspace{-4pt}
\begin{align}
\bar{\rho}^{\text{l}}<&\mbox{\small$\displaystyle\frac{\left(\!1\!+\!\mbox{erf}\!\left(\!\frac{0.5}{\sqrt{2}\sigma}\!\right)\!\!\right)^{\!K-1}}{2^K}\!\sum_{i=1}^N\!\bigg\{\!w_i
\mathbbm{1}(|\mathcal{Y}_i|\!=\!1)\!\!\left(\!1\!+\!\mbox{erf}\left(\!\frac{n_i\!-\!0.5}{\sqrt{2}\sigma}\!\right)\!\!\right)$}\notag\\
&+\textstyle w_i\mathbbm{1}(|\mathcal{Y}_i|\!>\!1)\left(\!1\!\!+\!\!\mbox{erf}\left(\frac{0.5}{\sqrt{2}\sigma}\right)\!\right)^{\!K\!-\!2}\!\!E_i\!\bigg\}.\label{eqn:local.Gau}
\end{align}
The plug-in estimate of the average expected DR-HA is\vspace{-3pt}
\begin{align} 
\!\bar{\hat{\rho}}^{\text{e}}\!<&\mbox{\small$\displaystyle
\!\frac{\left(\!1\!+\!\mbox{erf}\!\left(\!\frac{0.5}{\sqrt{2}\sigma}\!\right)\!\!\right)^{\!K-1}}{2^K}
\!\sum_{i=1}^N\!\bigg\{\!w_i\left(\!1\!+\!\mbox{erf}\left(\!\frac{n_i-0.5}{\sqrt{2}\sigma}\!\right)\!\!\right)\!\!\sum_{k=1}^{K}\!\hat{p}_{ik}^{n_i}\!\bigg\}+$}\notag\\
\!&\mbox{\small$\displaystyle\frac{\left(\!1\!\!+\!\!\mbox{erf}\left(\frac{0.5}{\sqrt{2}\sigma}\right)\!\right)^{\!K\!-\!2}}{2^K}\!\!\sum_{i=1}^N\!\bigg\{\!w_i\mathbbm{1}(n_i\!\ge\!2)\!\!\left(\!1\!-\!\sum_{k=1}^{K}\hat{p}_{ik}^{n_i}\!\right)\!E_i\!\bigg\}$}.\label{eqn:averagee.Gau.K}
\end{align}
Assume $\p_i\sim\mbox{Dirichlet}(\alpha_{1},\ldots,\alpha_{K})$ for $i=1,\ldots,N$, the average shrinkage DR-HA is
\begin{align}
\bar\rho^{\text{s}}\!<&\mbox{\small$\displaystyle\frac{\left(1\!+\!\mbox{erf}\!\left(\frac{0.5}{\sqrt{2}\sigma}\right)\!\right)^{\!K-1\!}\!}{2^{K}}\!\sum_{i=1}^{N}\bigg\{w_iA_i\left(1\!+\!\mbox{erf}\!\left(\frac{n_i-0.5}{\sqrt{2}\sigma}\right)\!\right)\bigg\}+$}\notag \\
&\mbox{\small$\displaystyle\frac{\left(\!1\!+\!\mbox{erf}\left(\!\frac{0.5}{\sqrt{2}\sigma}\!\right)\!\right)^{\!K-2}}{2^{K}}\!\!\sum_{i=1}^{N}\!\bigg\{w_i(1-A_i)\mathbbm{1}(n_i\!\ge\!2)E_i\bigg\}$}.\label{eqn:averages.Gau.K}
\end{align} 
Assume $n_i\!\sim\! f(n_i;\bs\beta)$ for $i\!=\!1,\ldots,N$, the plug-in estimate of the average marginal DR-HA is
\begin{align} 
\!&\!\bar{\hat{\rho}}^{\text{m}}\!\!<\!\mbox{$\small\displaystyle
\!\frac{\left(\!1\!+\!\mbox{erf}\!\left(\!\frac{0.5}{\sqrt{2}\sigma}\!\right)\!\!\right)^{\!K\!-\!1}}{2^K}\!\!
\sum_{i=1}^N\!\bigg\{\!w_if(n_i;\bs\beta)\! \left(\!1\!+\!\mbox{erf}\left(\!\frac{n_i\!-\!0.5}{\sqrt{2}\sigma}\!\right)\!\!\right)\!\!\sum_{k=1}^{K}\!\hat{p}_{ik}^{n_i}\!\bigg\}\!+\!\notag$}\\
\!&\mbox{$\small\displaystyle\frac{\left(\!1\!+\!\mbox{erf}\left(\frac{0.5}{\sqrt{2}\sigma}\right)\!\right)^{\!K\!-\!2}}{2^K}\!\!
\sum_{i=1}^N\!\!\bigg\{\!w_if(n_i;\bs\beta)\mathbbm{1}\!(n_i\!\ge\!2)\!\!\left(\!1\!-\!\!\sum_{k=1}^{K}\!\hat{p}_{ik}^{n_i}\!\right)\!\!E_i\!\bigg\}$},\!\!\!\label{eqn:averagem.Gau.K}
\end{align}
and the marginal shrinkage DR-HA is
\begin{align}
\rho^{\text{ms}}\!<&\mbox{$\small\displaystyle\frac{\!\left(1+\mbox{erf}\!\left(\frac{0.5}{\sqrt{2}\sigma}\right)\!\right)^{\!K-1\!}\!}{2^{K}}\!\!\sum_{n_i=1}^{\infty}\!\bigg\{\!f(n_i;\bs\beta)A_i\!\left(\!1\!+\!\mbox{erf}\!\left(\frac{n_i\!-\!0.5}{\sqrt{2}\sigma}\!\right)\!\right)\!\!\bigg\}$}\notag\\
+&\mbox{$\small\displaystyle\frac{\left(\!1+\!\mbox{erf}\left(\!\frac{0.5}{\sqrt{2}\sigma}\!\right)\!\right)^{\!K-2}}{2^{K}} \!\!\sum_{n_i=2}^{\infty}\left\{\!f(n_i;\bs\beta)(1-A_i)E_i\!\right\}$}.\!\!\label{eqn:averagems.gau}.
\end{align}
\end{thm}
The proof of Theorem \ref{thm:rho.gau.K}  is provided in the supplementary materials.  Parameters $\bs\beta$ and $\bs\alpha$ can be estimated in the same manner as in Theorem \ref{thm:rho.lap.K}, so is the specification of  $f(n_i;\bs\beta)$. 

For FDs  sanitized by Gaussian mechanisms, DR-HA  not only relates to $\epsilon$ but also  $\delta$. When $\epsilon$ or $\delta$ increases, $\sigma$ decreases, and DR-HA for cells in Scenario 1 increases; but the relationship is complicated for cells that fall in Scenario 8, due to similar reasons as in Theorem \ref{thm:rho.lap.K}. Also note that the Gaussian mechanism of $(\epsilon,\delta)$-DP requires $\epsilon<1$ whereas that of $(\epsilon,\delta)$-pDP does not impose any constraint on $\epsilon$, the latter would allow a more complete investigation of the relationship between DR-HA and ($\epsilon,\delta$) in this setting. 
Similar to Section \ref{sec:laplace}, we examine two special cases of Theorem \ref{thm:rho.gau.K} when $K=2$, the results of which are presented in the supplementary materials due to  space limitation, and when all original cells are homogeneous, the results of which are given in Corollary \ref{cor:rho.Gau.homo}. The proof of Corollary \ref {cor:rho.Gau.homo} is similar to that of  Corollary \ref{cor:rho.lap.homo}, so are the main conclusions.
\begin{cor}\label{cor:rho.Gau.homo}
When  all the original cells are homogeneous, the estimates of DR-HA measures in a sanitized FD via the Gaussian mechanisms  are
\begin{align}
\bar{\rho}^{\text{l}}&\!=\!\mbox{\small$\displaystyle\bar{\hat{\rho}}^{\text{e}}\!=\!
\frac{\left(1\!+\!\mbox{erf}\!\left(\!\frac{0.5}{\sqrt{2}\sigma}\!\right)\!\right)^{\!K-1}}{2^K}\!\sum_{i=1}^N\!\bigg\{\!w_i\!\left(\!1\!+\!\mbox{erf}\left(\!\frac{n_i\!-\!0.5}{\sqrt{2}\sigma}\!\right)\!\!\right)\!\!\!\bigg\}\!\!$}\label{eqn:averagel.Gau.homo}\\
\!\!\!\!\bar{\hat{\rho}}^{\text{s}}&=\mbox{\small$\displaystyle \frac{\left(\!1\!+\!\mbox{erf}\left(\!\frac{0.5}{\sqrt{2}\sigma}\!\right)\!\!\right)^{\!K\!-\!1\!}}{4 B(\alpha_1,\alpha_2)}\!\sum_{i=1}^N\!\bigg\{\!w_iA_i\left(\!1\!+\!\mbox{erf}\left(\!\frac{n_i-0.5}{\sqrt{2}\sigma}\!\right)\!\right)\!\!\bigg\}$}\label{eqn:averages.Gau.homo}\\
\!\!\!\!\bar{\rho}^{\text{m}}&\!=\!\mbox{\small$\displaystyle \!\frac{\left(\!1\!+\!\mbox{erf}\left(\!\frac{0.5}{\sqrt{2}\sigma}\!\right)\!\!\right)^{\!K\!-\!1\!}}{2^K}\!\!\sum_{n_i=1}^N\!\!\bigg\{w_if(n_i;\bs\beta)\!\left(\!1\!\!+\!\!\mbox{erf}\!\left(\!\frac{n_i\!-\!0.5}{\sqrt{2}\sigma}\!\right)\!\!\right)\!\!\!\bigg\}$}\!\!\!
\label{eqn:tilde.rho.Gau}\\
\!\!\!\!\rho^{\text{ms}}&=\frac{\!1\!+\!\mbox{erf}\!\left(\!\frac{0.5}{\sqrt{2}\sigma}\!\right)\!}{4B(\alpha_1,\alpha_2)}\!\!\sum_{n_i=1}^{\infty}\!\bigg\{\!f(n_i;\bs\beta)A_i\!\left(\!\!1\!+\!\mbox{erf}\left(\frac{n_i\!-\!0.5}{\sqrt{2}\sigma}\!\right)\!\!\right)\!\bigg\},\!\!\label{eqn:averagem.Gau.homo}
\end{align}
respectively. All four measures $\bar{\hat{\rho}}^{\text{e}},\bar{\rho}^{\text{s}}, \rho^{\text{m}},\hat{\tilde\rho}^{\text{m}}\in(2^{-K},1)$.
\end{cor}

\vspace{-9pt}\subsection{Extension to repeated FD publications} \label{sec:repeated} \vspace{-3pt}
Sections \ref{sec:laplace} and \ref{sec:gaussain} present the results on DR-HA for a single sensitive attribute $Y$ in a sanitized FD. When there are multiple $Y$'s, the DR-HA when releasing each $Y$ can be evaluated separately, applying the results in Sections \ref{sec:laplace} and \ref{sec:gaussain}. If there is a need to aggregate DR-HA over multiple $Y$'s, one would first define what overall DR-HA is. For example, for a given DR-HA measure $\rho$ in Section \ref{sec:DR-HA}, the overall DR-HA in a cell can be defined as $\max_j\rho_j$, referred to as the \emph{bounded measure}, or  as $\sum_j\rho_j$, referred to as the  \emph{unbounded measure}, where $\rho_j$ is the DR-HA for $Y_j$ in that cell for $j=1,\ldots,q$.

After the overall DR-HA is defined, one may examine its relationship with privacy loss, which depends on how FDs are sanitized and released.  If the  FD formed by $\X$ and all $Y$'s is sanitized and released once given a privacy loss parameter, i.e., not repeated publication, we may calculate the DR-HA for each $Y$ separately, aggregate across the $Y$'s to obtain an overall DR-HA, and vary the privacy loss to examine how the overall DR-HA changes. The more interesting case is when the $Y$'s are released sequentially, which is repeated publication.  Since all the $Y$'s share the same set of QIDs $\X$, each additional release implies further splitting of the cells formed by $\X$ and the $Y$'s released earlier. The counts in the newly generated children cells are sanitized, under the equality constraints imposed by the previously released  parent cell counts (summation of children cell counts equal to their parent cell count). To calculate the overall DR-HA, one may first apply the formulas  in Sections \ref{sec:laplace} and \ref{sec:gaussain} to each $Y$  at each release at the most updated cumulative privacy loss.  Suppose we spend $(\epsilon_0,\delta_0)$ on sanitizing  FD$(\X)$ and $(\epsilon_j,\delta_j)$ on sanitizing the cells in  FD$(\X,Y_1,\ldots,Y_j)$ for $j\!=\!1,\ldots, q$. The DR-HA from releasing $Y_j$ would be evaluated at privacy loss $(\epsilon_0\circ\epsilon_1\circ\ldots\circ\epsilon_j,\delta_0\circ\delta_1\circ\ldots\circ\delta_j)$, where $\circ$ stands for a privacy loss composition operator that can be the basic sequential composition, the advanced composition, or others.\footnote{$(\epsilon_0\circ\epsilon_1\circ\ldots\circ\epsilon_j,\delta_0\circ\delta_1\circ\ldots\circ\delta_j)$ is an upper bound for the actual privacy loss associated with  FD$(\X,Y_j)$ for $j\ge2$ the exact value of which lies between $(\epsilon_0\circ\epsilon_j,\delta_0\circ\delta_j)$ and $(\epsilon_0\circ\epsilon_1\circ\ldots\circ\epsilon_j, \delta_0\circ\delta_1\circ\ldots\circ\delta_j)$.} Finally, one may evaluate the overall DR-HA in releasing sanitized FD$(\tilde{\X},\tilde{Y}_1,\ldots,\tilde{Y}_j)$ and examine how it changes with the cumulative privacy loss $(\epsilon_0\circ\epsilon_1\circ\ldots\circ\epsilon_j, \delta_0\circ\delta_1\circ\ldots\circ\delta_j)$. For the bounded overall DR-HA measure, once it reaches the maximum after a certain release, it will stay there in any future release; for the unbounded DR-HA, it will monotonically increase in $j$.

The above can be generalized to cases where new information is constantly collected and added to a database and new queries are requested from the data. The overall DR-HA will change during the process, so will its relationships with the cumulative privacy loss,  depending what new information is collected and what additional FD queries are released.

First, there is horizontal growth in the data after the release of FD$(\tilde{\X},\tilde{\Y})$. We define horizontal growth as new information $\Z$ being collected from the same set of individuals previously released. $\Z$ may contain brand new attributes or repeated measures of released attributes. Adding $\Z$ to FD$(\tilde{\X},\tilde{\Y})$ would lead to further splitting of the released cells. If $\Z$ contains sensitive attributes and no identifiable information, then the analysis above on the sequential release of FD$(\tilde{\X},\tilde{Y}_1,\ldots,\tilde{Y}_j)$ applies as $\Z$ is essentially just new $Y$'s. If $\Z$  contains only PIDs and no sensitive attributes, each additional data publication would lead to higher DR-HA on the sensitive attributes.  As the number of PIDs increases,  the number of homogeneous cells increases\footnote{Homogeneous cells released at $t_0$ remains homogeneous and heterogeneous cells at $t_0$ would get a chance to become homogeneous after being split further by new PIDs}, and the cell sizes decrease, leading to higher DR-HA. In terms of the DR-HA calculation, the  formulas in Sections \ref{sec:laplace} and \ref{sec:gaussain} still apply -- at the most updated cumulative privacy loss in the newly constructed FD$((\X,\Z),\Y)$. 
If $\Z$ contains both PIDs $\Z_X$ and  sensitive attributes $\Z_Y$, the steps of updating DR-HA would be a combination of the above two cases. Specifically, one would first apply the formulas  to calculate the DP-HA for each sensitive attribute in $(\Y,\Z_Y)$ in the cells of the $(\X,\Z_X)$  cross-tabulation at the updated privacy loss, and then evaluate the overall DR-HA from releasing FD$(\tilde{\X},\tilde{\Y},\tilde{\Z})$ .

Second, there is vertical grow in the data after the release of FD$(\tilde{\X},\tilde{\Y})$, meaning that the set of attributes remain the same and information is collected on these attributes for a set of new cases. Assuming that the cases in the new cohort have no overlapping information with the cases released previously, the privacy loss does not accumulate with additional releases per the parallel privacy loss composability and it is thus not meaningful to measure the overall DR-HA across the cohorts,  and one may evaluate DR-HA separately for each cohort with the formulas in Sections \ref{sec:laplace} and \ref{sec:gaussain} at the updated privacy loss in each cohort. 

The most complicated scenario is that a new publication occurs when the newly collected information is a mixture of vertical and horizontal growth since the last release, or only horizontal growth in a subset of released records. In the former, one would update the DR-HA for the previously released cohort given the horizontal growth at the cumulative privacy loss and for the new cohort at its privacy loss separately. In the latter case, only the subset with the horizontal growth needs to be updated on DR-HA, which may go up drastically  given that it is a smaller cohort. 
\begin{figure*}[!tb]
\vspace{-15pt}\centering
\subfloat[Adult]{{\includegraphics[width=0.375\textwidth, height=0.175\textwidth]{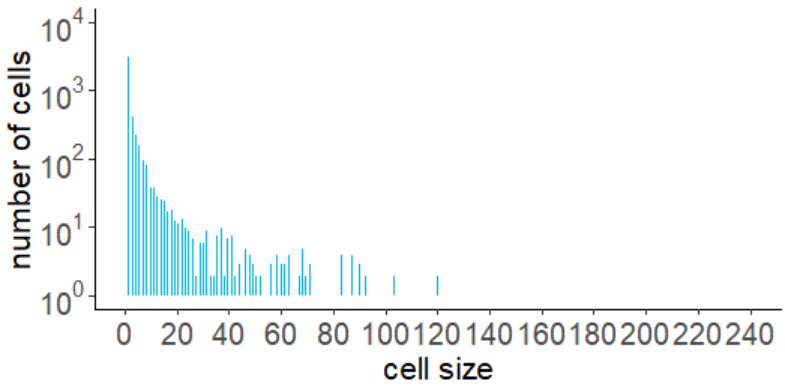}}}
\vspace{-6pt}
\subfloat[Bankruptcy]{{\includegraphics[width=0.24\textwidth]{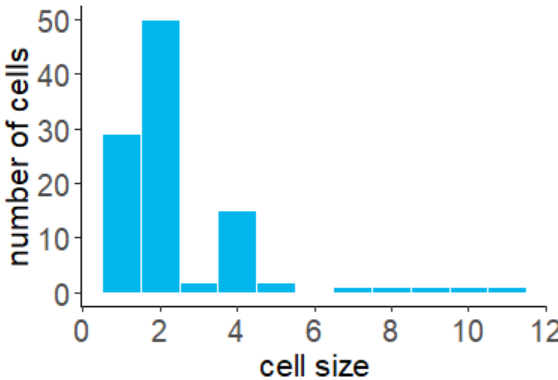}}}
\subfloat[Bankruptcy subset]{{\includegraphics[width=0.245\textwidth]{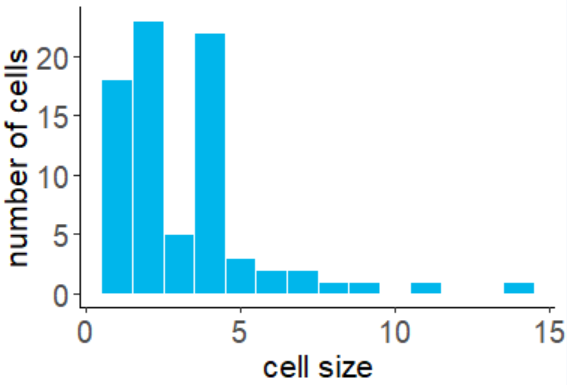}}}
\caption{Histograms of cell size $n_i$ in the experiments ($n_i\in[1, 240]$ in the Adult data and $\#\{n_i>120\}=6$).} \label{fig:hist}\vspace{-12pt}
\end{figure*}

\normalsize\vspace{-9pt}
\section{Experiments} \label{sec:experiment}\vspace{-3pt}
We apply the derived mathematical relations between DR-HA and privacy loss parameters  in Section \ref{sec:relation} to 3 experiments using the Adult data \citep{Adult1996},  the Bankruptcy data \citep{Bankruptcy2014}\footnote{Both datasets are  available at the UCI Machine Learning Repository.}, and a subset of the Bankruptcy data.  We demonstrate how to make use of the relations to choose $\epsilon$ and $\delta$. In all the experiments, we examine $\epsilon\in[10^{-3},10^2]$ for $\epsilon$-DP and $(\epsilon,\delta)$-pDP, $\epsilon\in[10^{-3},1]$ for $(\epsilon,\delta)$-DP, and $\delta\in(O(n^{-1}), 0.1]$, where  $n$ is the data sample size, $n^{-1}=3.6\times 10^{-5}$ for the Adult data and $4\times 10^{-3}$ for the Bankruptcy data and the subset data. The larger values of $\epsilon$ and $\delta$ are for scientific investigation only and unlikely to be used in practical applications.

We present the main observations from the experiments below; the experiment settings and detailed results are given in Sections \ref{sec:setting} and \ref{sec:results}. 
\begin{enumerate}[leftmargin=12pt, itemsep=-3pt]
\item DP sanitization of FDs mitigates DR-HA. 
\item The relationship between DR-HA and log($\epsilon$) for the Laplace mechanism and the Gaussian  mechanism of $(\epsilon,\delta)$-pDP follows an S shape. DR-HA is close to the minimum for $\epsilon\!<\!0.1$ and reasonably small for  $\epsilon\!<\!1$. The drastic increase in DR-HA occurs when $\epsilon\!\in\!(1,10)$, and reaches the maximum for $\epsilon\!>\!10$.  Comparatively, the impact of $\delta$ on DR-HA is relatively insignificant.
\item The relationships between  DR-HA and $\epsilon$ in the Adult and Bankruptcy experiments with 100\% homogeneous cells are similar although they have different $n,N,$ and $\n_i$. 
\item $\epsilon$ around 1 seems to be a good choice from the perspective of protecting against DR-HA, at least in data similar to the three experiments. $O(n^{-1})$ is a good choice for $\delta$ given its negligible impact on DR-HA and  smaller impact on utility compared to $\epsilon$. We recommend fixing $\delta$ first and choosing $\epsilon$ when Gaussian mechanisms are used.
\end{enumerate}
\begin{figure*}[!tb] 
Adult data  \hspace{0.7in}
(a) - (c): $\bar{\rho}^{\text{l}}_{\text{uw}} = \bar{\hat{\rho}}^{\text{e}}_{\text{uw}}$\hspace{2.5in}
(d) - (f):  $\bar{\rho}^{\text{l}}_{\text{w}} = \bar{\hat{\rho}}^{\text{e}}_{\text{w}}$ \vspace{-9pt}\\
\subfloat[$\epsilon$-DP] 
{\includegraphics[scale=0.315]{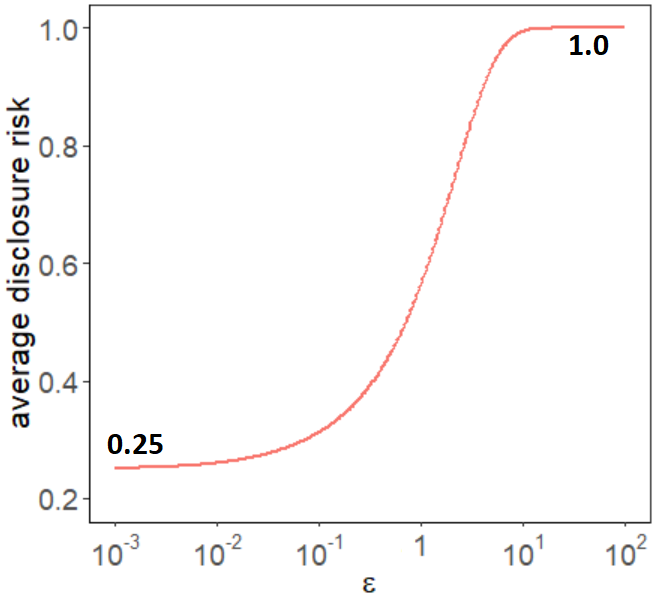}}
\subfloat[$(\epsilon,\delta)$-DP] {\includegraphics[scale=0.285]{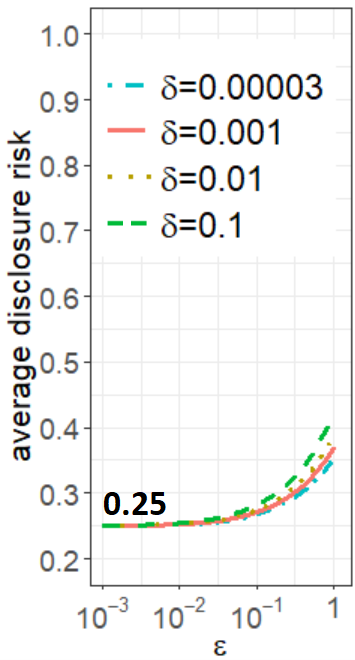}}
\subfloat[$(\epsilon,\delta)$-pDP] {\includegraphics[scale=0.285]{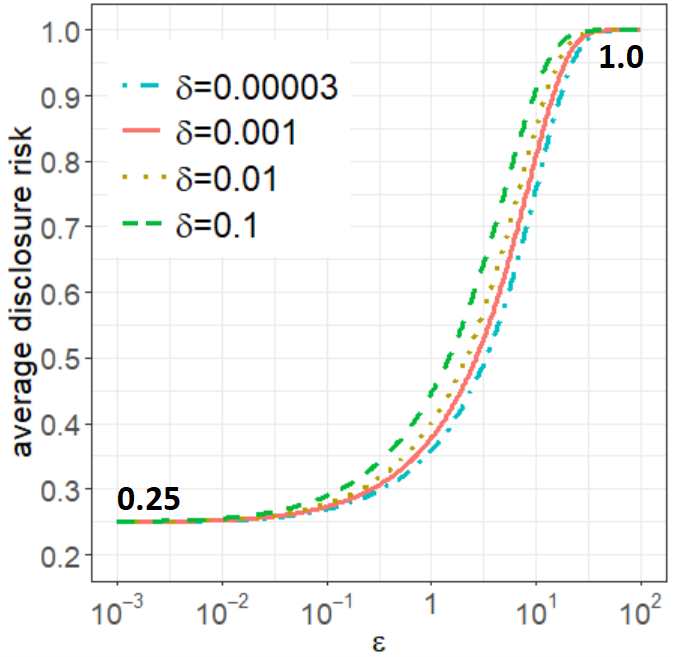}}
\subfloat[$\epsilon$-DP] 
{\includegraphics[scale=0.315]{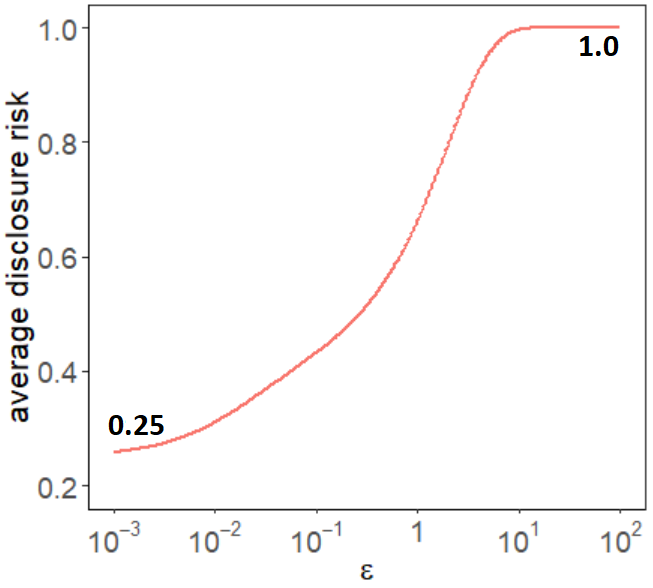}}
\subfloat[$(\epsilon,\delta)$-DP] {\includegraphics[scale=0.315]{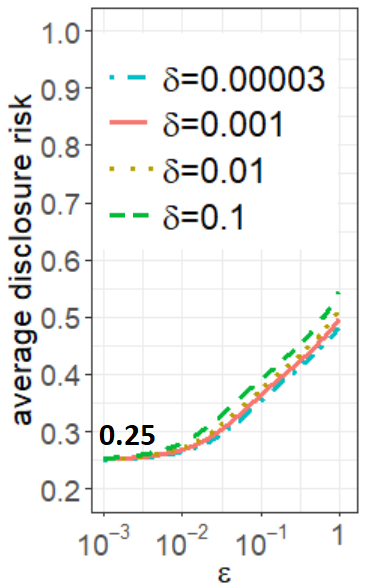}}
\subfloat[$(\epsilon,\delta)$-pDP] {\includegraphics[scale=0.315]{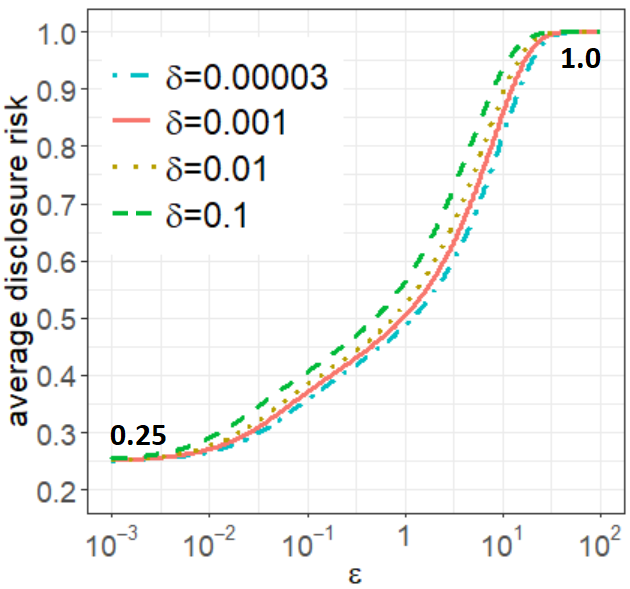}}\\
\vspace{6pt} Bankruptcy data: \hspace{0.6in}
(g)- (i): $\bar{\rho}^{\text{l}}_{\text{uw}} = \bar{\hat{\rho}}^{\text{e}}_{\text{uw}}$\hspace{2.5in}
(j) - (l) $\bar{\rho}^{\text{l}}_{\text{w}} = \bar{\hat{\rho}}^{\text{e}}_{\text{w}}$\vspace{-12pt}\\
\subfloat[$\epsilon$-DP] 
{\includegraphics[scale=0.315]{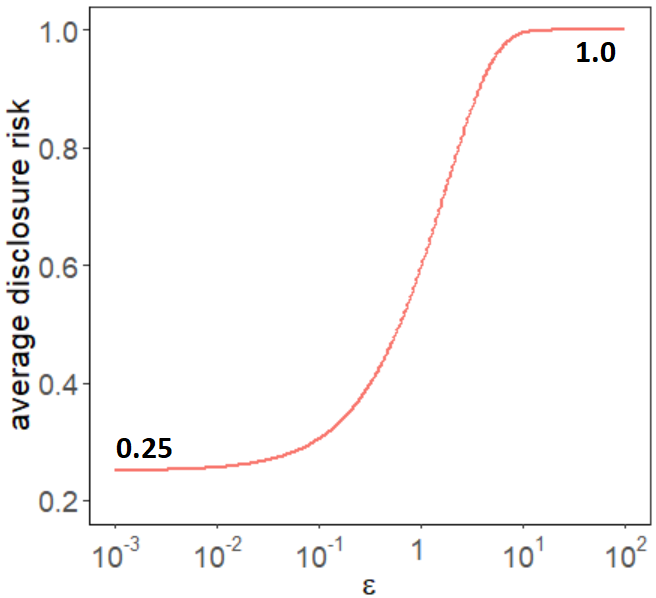}}
\subfloat[$(\epsilon,\delta)$-DP] {\includegraphics[scale=0.285]{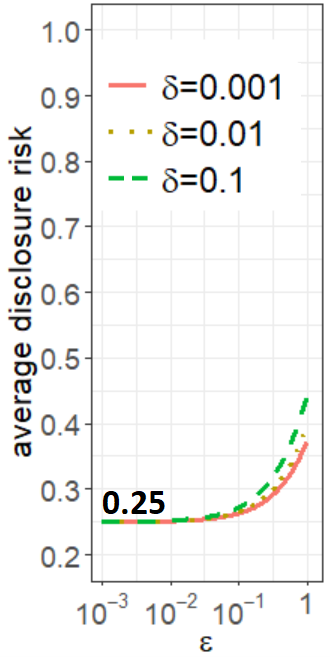}}
\subfloat[$(\epsilon,\delta)$-pDP] {\includegraphics[scale=0.285]{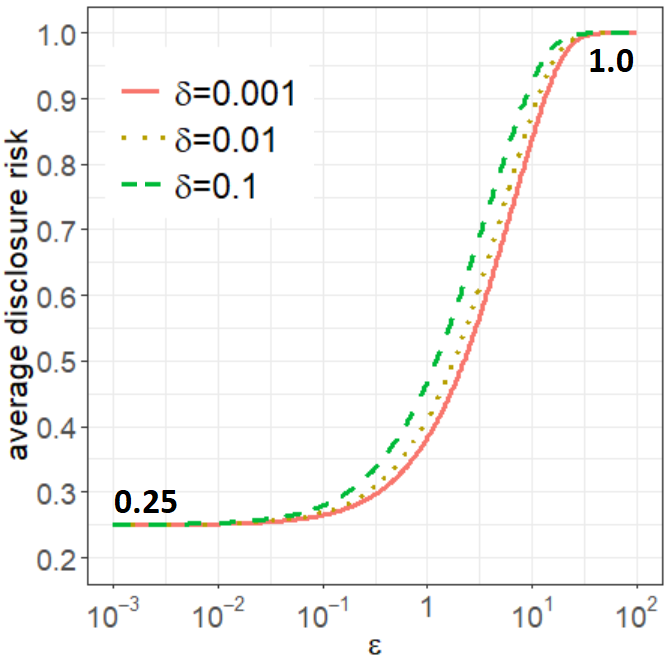}}
\subfloat[$\epsilon$-DP] 
{\includegraphics[scale=0.315]{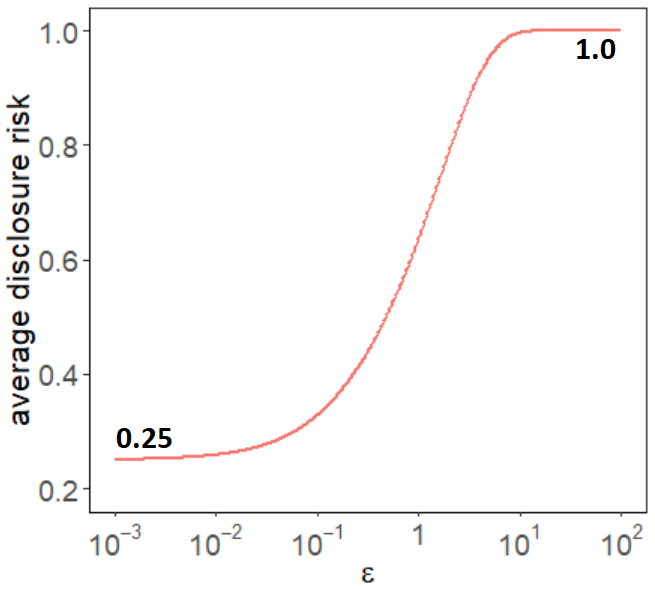}}
\subfloat[$(\epsilon,\delta)$-DP] {\includegraphics[scale=0.315]{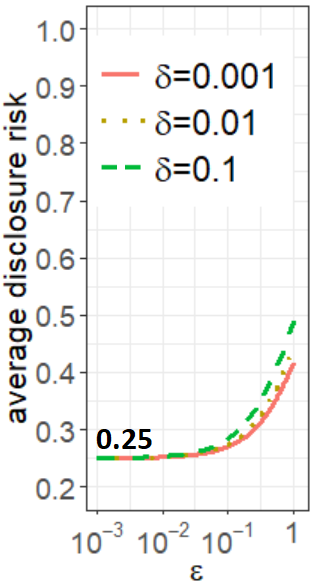}}
\subfloat[$(\epsilon,\delta)$-pDP] {\includegraphics[scale=0.32]{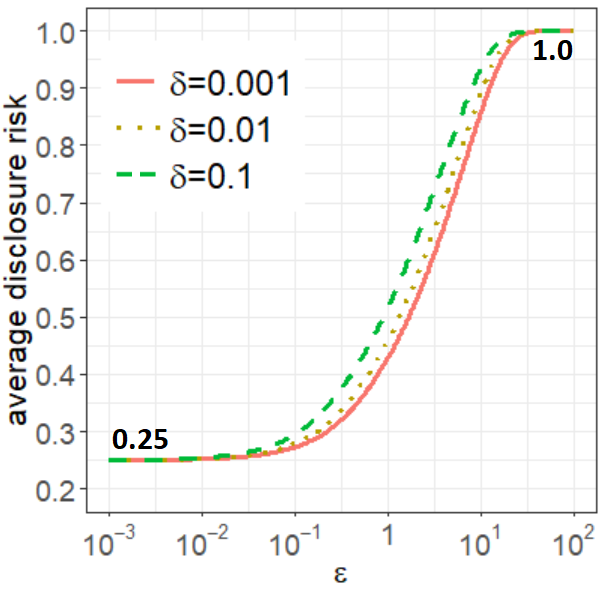}}
\vspace{3pt}
Bankruptcy subset: \hspace{0.6in}
(m) - (o): $\bar{\rho}^{\text{l}}_{\text{uw}}$ \hspace{2.5in}
(p) - (r): $\bar{\rho}^{\text{l}}_{\text{w}}$ \vspace{-9pt}\\
\subfloat[$\epsilon$-DP] {\includegraphics[scale=0.37]{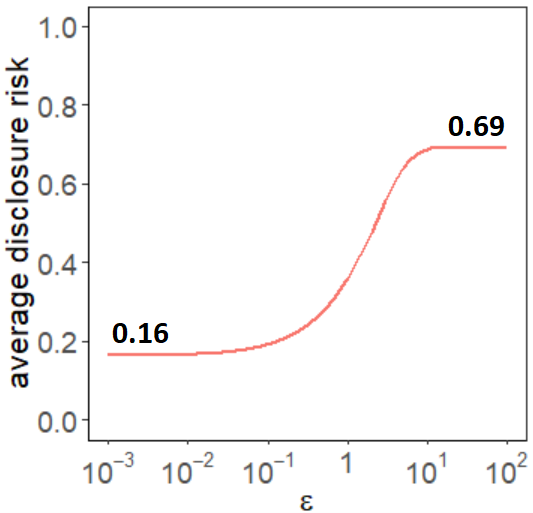}}
\subfloat[$(\epsilon,\delta)$-DP] {\includegraphics[scale=0.36]{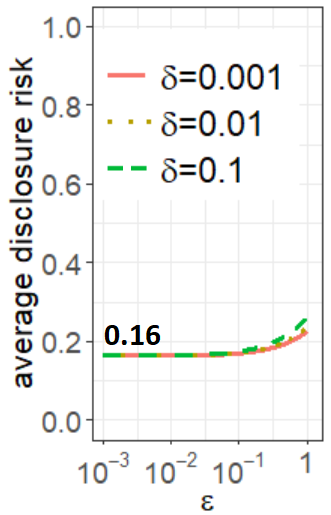}}
\subfloat[$(\epsilon,\delta)$-pDP] {\includegraphics[scale=0.36]{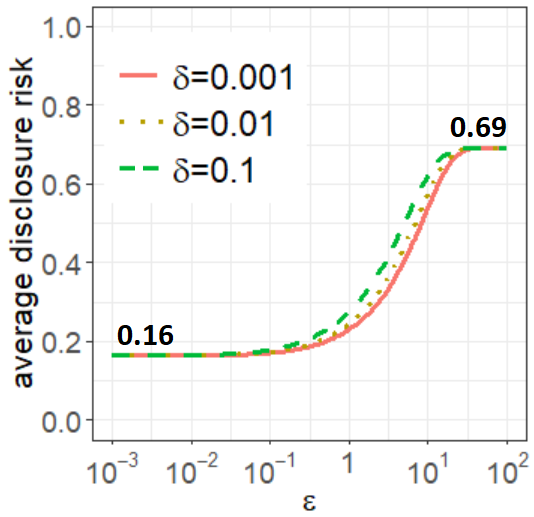}}
\subfloat[$\epsilon$-DP]
{\includegraphics[scale=0.32]{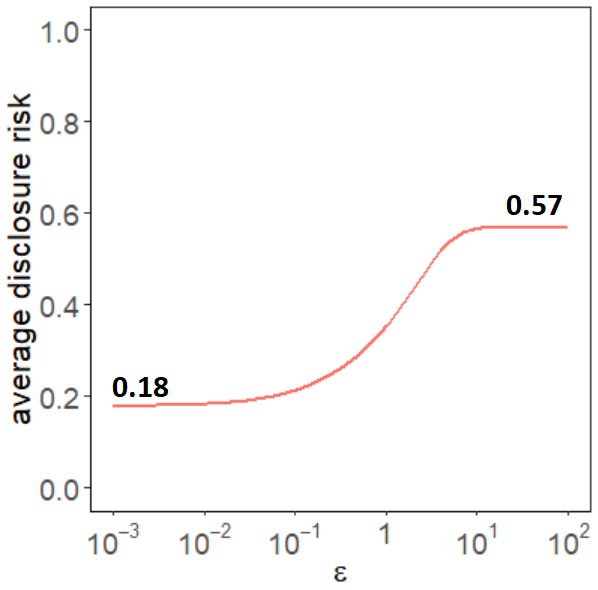}}
\subfloat[$(\epsilon,\delta)$-DP] {\includegraphics[scale=0.32]{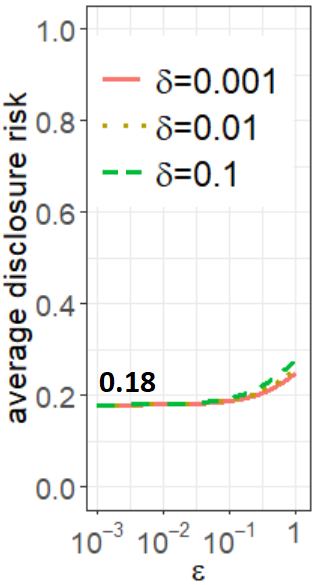}}
\subfloat[$(\epsilon,\delta)$-pDP] {\includegraphics[scale=0.32]{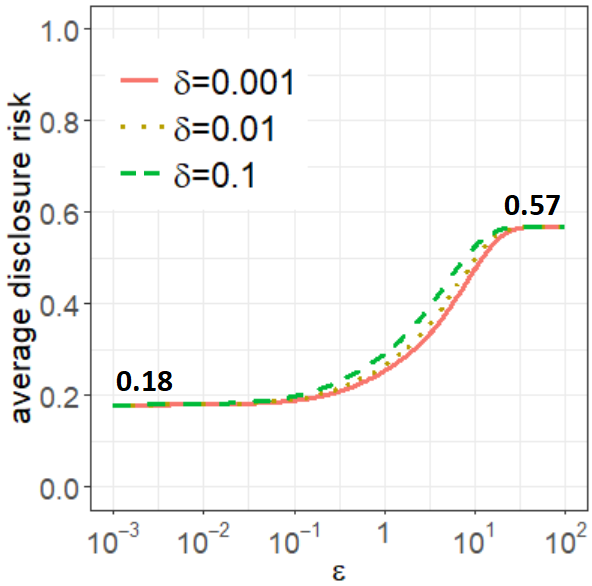}}
\vspace{3pt}
Bankruptcy subset: \hspace{0.6in}
(s) - (u): $\bar{\hat{\rho}}^{\text{e}}_{\text{uw}}$ \hspace{2.5in}
(v) - (x): $\bar{\hat{\rho}}^{\text{e}}_{\text{w}}$\vspace{-9pt}\\
\subfloat[$\epsilon$-DP]
{\includegraphics[scale=0.305]{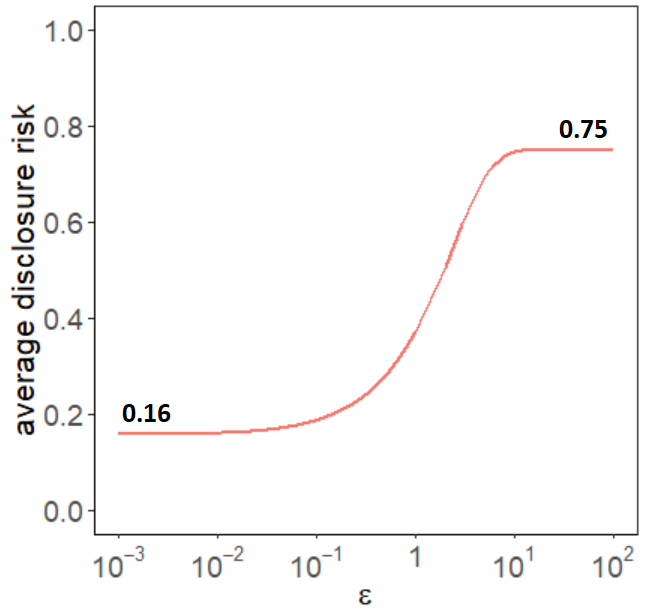}}
\subfloat[$(\epsilon,\delta)$-DP] {\includegraphics[scale=0.30]{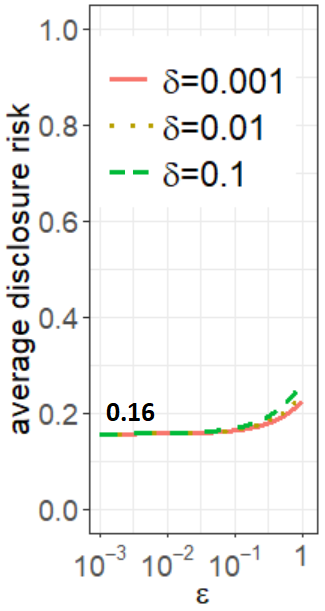} }
\subfloat[$(\epsilon,\delta)$-pDP] {\includegraphics[scale=0.30]{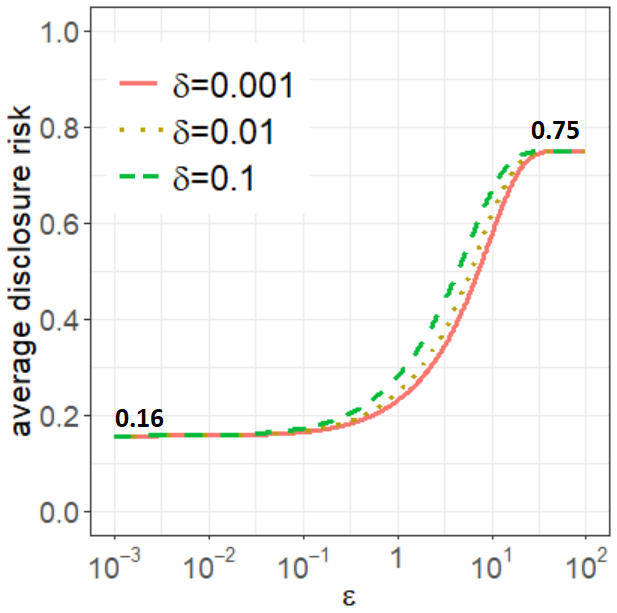}}
\subfloat[$\epsilon$-DP] {\includegraphics[scale=0.32]{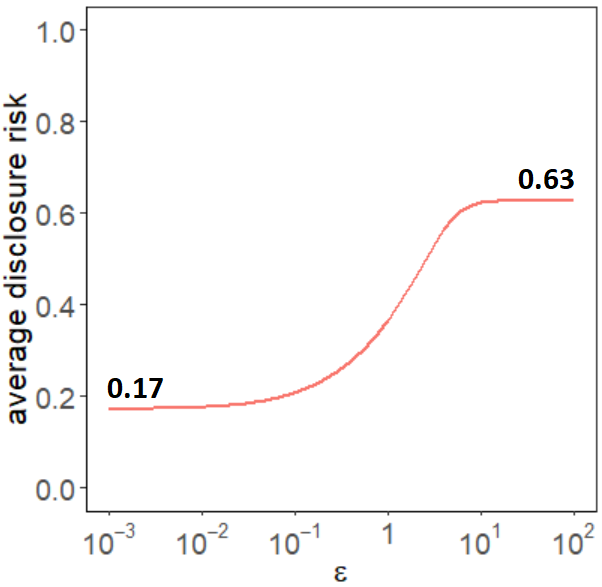}}
\subfloat[$(\epsilon,\delta)$-DP] {\includegraphics[scale=0.32]{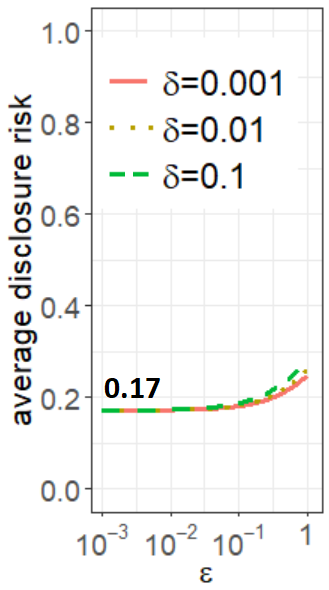}}
\subfloat[$(\epsilon,\delta)$-pDP] {\includegraphics[scale=0.32]{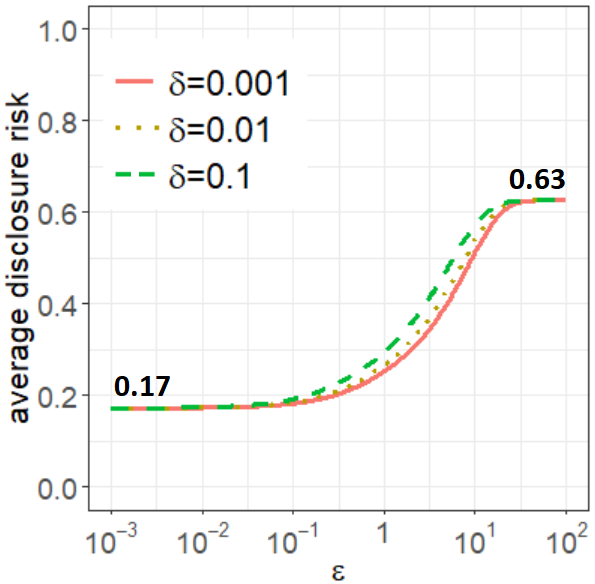}}
\caption{Relationships between $(\epsilon,\delta)$ and  weighted and unweighted DR-HA measures $\bar{\rho}^{\text{l}}$  and $\bar{\hat{\rho}}^{\text{e}}$ in sanitized FDs (Gaussian mechanism of $(\epsilon,\delta)$-DP requires $\epsilon<1$ and thus the partial curves in (b), (e), (h), (k), (n), (q), (t) and (w)).} \label{fig:rhole}\vspace{-12pt}
\end{figure*}

\vspace{-9pt}\subsection{Experiment setting}\label{sec:setting}\vspace{-4pt}
The Adult data contain 27,504 individuals. We treat Age, Relationship, Education, Race, Sex, Hours-per-week (6 attributes) as QIDs and the binary Income as the sensitive attribute ($\le\!50K$, $>\!50K$). The FD over the 6 QIDs (Age is grouped every $5$ years and Hours-per-week is grouped every $10$ hours) is a 6-dimensional histogram with $N\!=\!5,974$ non-empty cells and each cell is homogeneous on income.   The Bankruptcy dataset contains 250 observations and 7 categorical attributes. We treat the 6 qualitative variables (Credibility/Cr, Competitiveness/Co, Financial Flexibility/FF, Industrial Risk/IR, Management Risk/MR, and Operating Risk/OR) as QIDs, each of which has 3 levels (Positive, Average, Negative), and Bankruptcy status (Bankruptcy, Non-Bankruptcy) as a sensitive attribute. The cross-tabulation of the 6 QIDs leads to $N\!=\!103$ non-empty homogeneous cells on the Bankruptcy status.  To demonstrate the relationships in data with heterogeneous cells and $K\!>\!2$, we also use a subset of the attributes of the Bankruptcy data ($n$ is still 250) to create an FD with heterogeneous cells and $K\!=\!3$. Specifically, we treat IR, MR, Cr, Co, and OR as QIDs, and FF as the sensitive attribute ($K\!=\!3$). The QIDs lead to 78 non-empty cells, among which 54 are homogeneous and 24 are heterogeneous. The distribution of the cell sizes  ($n_i$) in each experiment is depicted in Figure \ref{fig:hist}.

For demonstration purposes, we calculate the relationships between $\bar{\rho}^{\text{l}},\bar{\hat{\rho}}^{\text{e}}$  and ($\epsilon,\delta$).\footnote{The relations between $\bar{\hat{\rho}}^{\text{m}}, \bar{\hat{\rho}}^{\text{s}},\hat{\rho}^{\text{sm}}$  and ($\epsilon,\delta$)  can also be obtained using the results from Section \ref{sec:relation}, assuming distributions on $\mathbf{p}$ and $n_i$. For example, if $n_i\sim$ Pois($\beta$), the ML estimate of $\beta$ is 4.6 for the Adult data, 2.43 for the Bankruptcy data, and 3.21 for the Bankruptcy subset).} In the Adult and Bankruptcy data with 100\%  homogeneous cells, we applied  Corollaries \ref{cor:rho.lap.homo} and \ref{cor:rho.Gau.homo} at $K=2$ to calculate $\bar{\rho}^{\text{l}}$ and $\bar{\hat{\rho}}^{\text{e}}$. In the  subset Bankruptcy data with heterogeneous cells, we applied Eqns  (\ref{eqn:averagee.lap.K}) and (\ref{eqn:averagee.Gau.K}) in Theorems \ref{thm:rho.lap.K} and \ref{thm:rho.gau.K} to calculate $\bar{\rho}^{\text{l}}$ and $\bar{\hat{\rho}}^{\text{e}}$.\footnote{The results are obtained based on the formulas from Sec  \ref{sec:relation}, and no actual sanitization is needed, which is one of the motivations for us to derive the formulas in the first place. We did calculated the empirical DR-HA, by sanitizing  the FD in the cross-tabulation of QIDs $\X$ and $Y$ via the Laplace and Gaussian mechanism at the examined values of $\epsilon$ and $(\epsilon,\delta)$; the empirical results are nearly identical to the theoretical results, except for Monte Carlo errors in the former.}

\vspace{-9pt} \subsection{Relationships between $(\epsilon,\delta)$ and DR-HA}\label{sec:results}\vspace{-4pt}
The results are presented in Figure \ref{fig:rhole}. Note that the relations of $\bar{\rho}^{\text{l}}$ vs log($\epsilon$) and $\bar{\hat{\rho}}^{\text{e}}$ vs log($\epsilon$) are the same when all the original cells are homogeneous (the Adult and Bankruptcy data) per Corollaries \ref{cor:rho.lap.homo} and \ref{cor:rho.Gau.homo}.  In summary,  (1) all relationships are S-shaped except for the Gaussian mechanisms of $(\epsilon,\delta)$-DP that requires $\epsilon<1$. (2)  In the Adult and Bankruptcy  experiments with 100\% homogeneous cells and $K\!=\!2$, the upper asymptote of DR-HA is 1 (the DR-HA value in the original data) when $\epsilon\!\ge\!10$ for the Laplace  mechanism and  when $\epsilon\!\ge\sim\!31.6$ for the Gaussian mechanism of $(\epsilon,\delta)$-pDP regardless of  $\delta$. The lower asymptote is $0.25$ ($2^{-K}$ per Corollaries \ref{cor:rho.lap.homo} and \ref{cor:rho.Gau.homo}) regardless of $\epsilon$ or $\delta$. (3) In the subset Bankruptcy data with  heterogeneous cells and $K\!=\!3$, the upper asymptote of $\bar{\rho}^{\text{l}}_{\text{uw}}, \bar{\rho}^{\text{l}}_{\text{w}}, \bar{\hat{\rho}}^{\text{e}}_{\text{uw}}$,  and $\bar{\hat{\rho}}^{\text{e}}_{\text{w}}$ is 0.69, 0.57, 0.75, and 0.63, respectively. Specifically, $\bar{\rho}^{\text{l}}_{\text{uw}}$ approaches the proportion of homogeneous cells (54 out of 78) as $\epsilon\rightarrow\infty$; the $\bar{\rho}^{\text{l}}_{\text{w}}$ approaches the proportion of individuals in the homogeneous cells (142 out of 250), whereas $\bar{\hat{\rho}}^{\text{e}}_{\text{uw}}$ and $\bar{\hat{\rho}}^{\text{e}}_{\text{w}}$  integrate out the sampling error around $\n_i$ in each cell, measure the expected DR-HA for any dataset that has the same underlying distribution as $\n_i$ and sanitized by the same $\M$, and converge to $N^{-1} \sum_i (\hat{p}_i^{n_i}+(1-\hat{p}_i)^{n_i})$ and $\sum_i\big((n_i/n) (\hat{p}_i^{n_i}+(1-\hat{p}_i)^{n_i})\big)$, respectively, as $\epsilon\rightarrow\infty$. The lower asymptote is the same for $\bar{\rho}^{\text{l}}_{\text{uw}}$ and $\bar{\hat{\rho}}^{\text{e}}_{\text{uw}}$, which is 0.16 and similar to that for  $\bar{\rho}^{\text{l}}_{\text{w}}$ (0.18) and $\bar{\hat{\rho}}^{\text{e}}_{\text{w}}$ (0.17).  (4) Relative to the impact of $\epsilon$ on DR-HA, the impact of $\delta$ on DR-HA is relatively minor in the Gaussian mechanisms. (4) Between the two Gaussian mechanisms, DR-HA for the Gaussian mechanism of $(\epsilon,\delta)$-pDP is slightly larger ($+0.05$) than that of $(\epsilon,\delta)$-DP for $\epsilon<1$.  

\vspace{-6pt} 
\subsection{Choosing $(\epsilon,\delta)$, balancing DR-HA and Utility } \label{sec:utility}\vspace{-3pt}
To demonstrate how one may consider both DR-HA and sanitized data utility to choose privacy loss parameters when sanitizing information, we release 1-way, 2-way and 3-way marginals from the subset Bankruptcy data as an example. 
For the utility analysis, we calculated the total variation distance (TVD) between the original and sanitized probability distributions in the marginals.\footnote{TVD $=||\hat{\mathbf{p}}-\hat{\tilde{\mathbf{p}}}||_1/2$, where $\hat{\mathbf{p}}$ and  $\hat{\tilde{\mathbf{p}}}$ are the sample probabilities of the cells in a cross-tabulation based on the original and sanitized data, respectively.  For example, Cr has 3 categories, the original sample probabilities are $\hat{\mathbf{p}}\!=\!(p_1,p_2,p_3)$  and the sanitized probabilities are $\hat{\tilde{\mathbf{p}}}\!=\!(\hat{\tilde{p}}_1,\hat{\tilde{p}}_2,\hat{\tilde{p}}_3)$; the cross-tabulation of Cr and Co results in 9-cell 2-way marginals with original cell probabilities $\hat{\mathbf{p}}\!=\!(\hat{p}_{11},\ldots,\hat{p}_{33})$ and sanitized probabilities $\hat{\tilde{\mathbf{p}}}\!=\!(\hat{\tilde{p}}_{11},\ldots,\hat{\tilde{p}}_{33})$.}  In total, there are 6 1-way marginals, 15 2-way marginals, and 20 3-way marginals. 

\begin{figure}[!htb]
\vspace{-3pt} \centering
{\includegraphics[width=0.35\linewidth, height=0.32\linewidth] {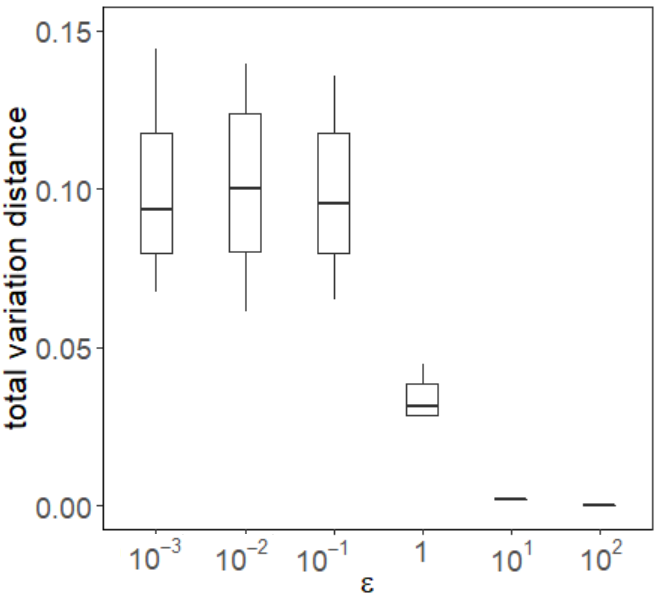}}
{\includegraphics[width=0.1\textwidth, height=0.32\linewidth] {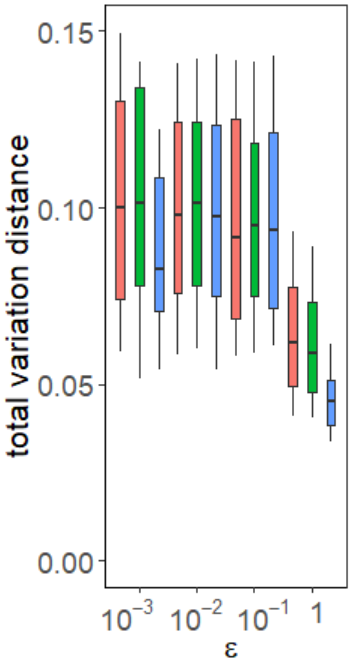}}
{\includegraphics[width=0.18\textwidth, height=0.32\linewidth] {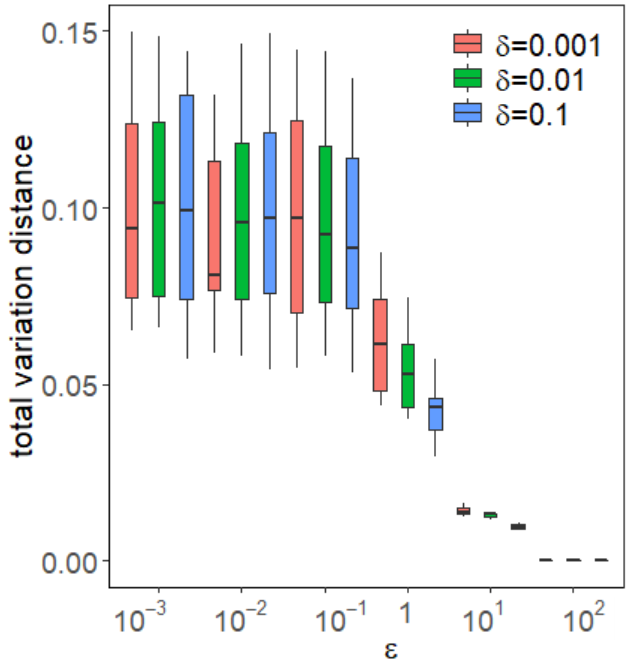}}\\ 
\footnotesize{1-way TVD: $\epsilon$-DP \hspace{0.5cm}  $(\epsilon,\delta)$-DP  \hspace{1.2cm} $(\epsilon,\delta)$-pDP}\\
{\includegraphics[width=0.35\linewidth, height=0.32\linewidth] {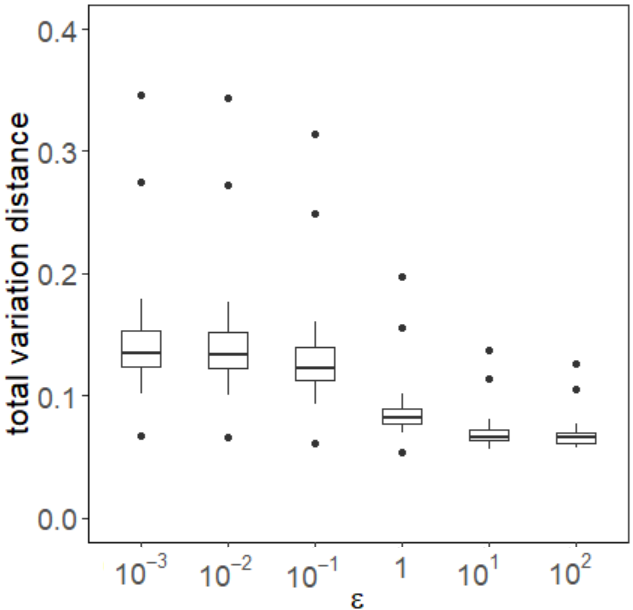}}
{\includegraphics[width=0.1\textwidth, height=0.32\linewidth] {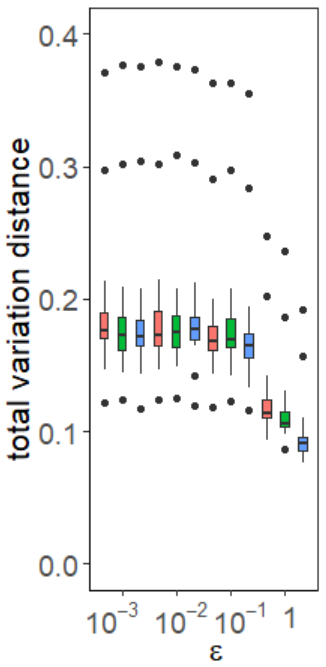}}
{\includegraphics[width=0.18\textwidth, height=0.32\linewidth] {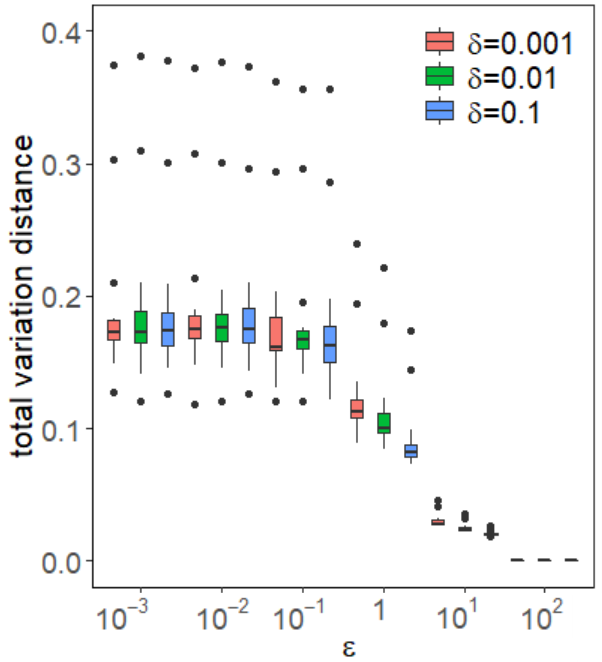}}\\ 
\footnotesize{2-way TVD: $\epsilon$-DP \hspace{0.5cm}  $(\epsilon,\delta)$-DP  \hspace{1.2cm} $(\epsilon,\delta)$-pDP}\\
{\includegraphics[width=0.35\linewidth, height=0.32\linewidth] {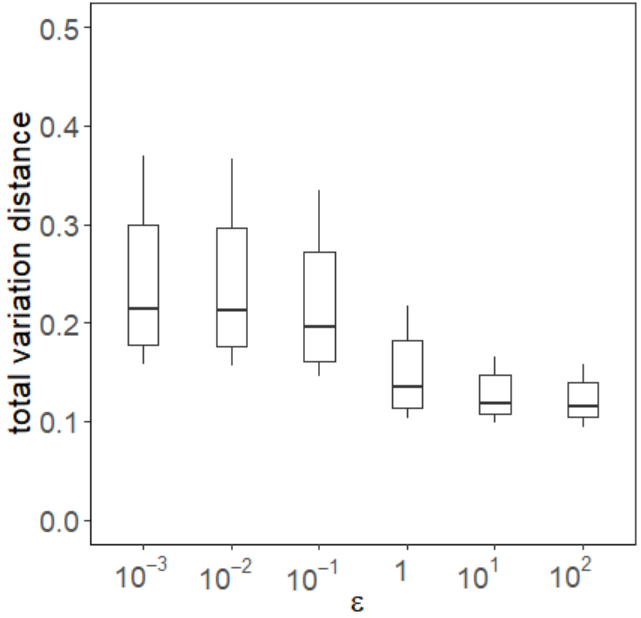}}
{\includegraphics[width=0.1\textwidth, height=0.32\linewidth] {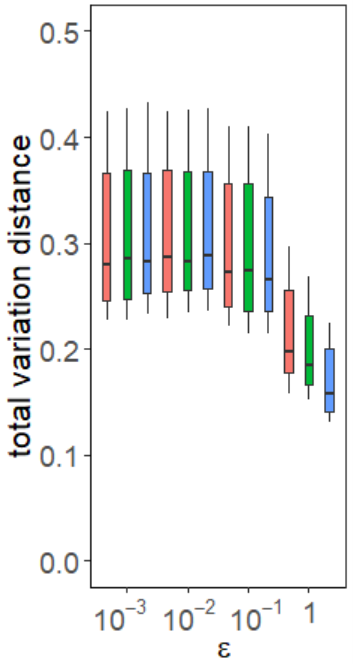} }
{\includegraphics[width=0.18\textwidth, height=0.32\linewidth] {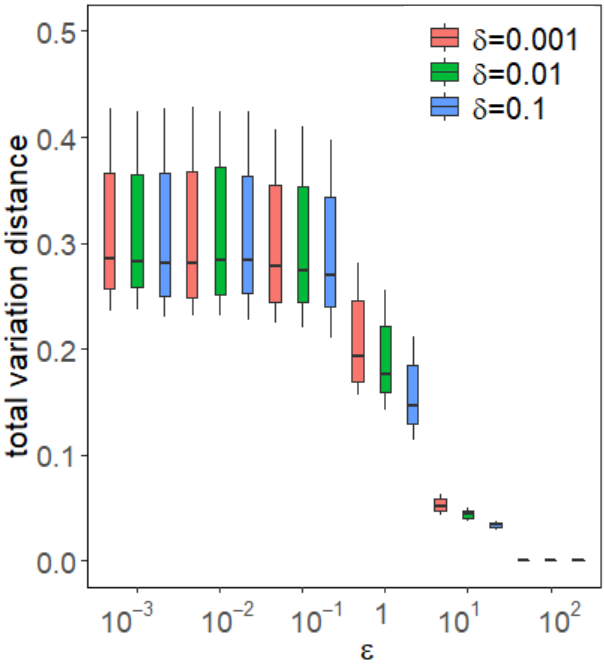}}\\ 
\footnotesize{3-way TVD: $\epsilon$-DP \hspace{0.5cm}  $(\epsilon,\delta)$-DP  \hspace{1.2cm} $(\epsilon,\delta)$-pDP}
\vspace{-6pt} 
\caption{Total variation distance between original and sanitized sample probability distributions in 1-way, 2-way and 3-way marginals.}\label{fig:utility} \vspace{-12pt} 
\end{figure}

The box plots of the TVD of the marginals at various $\epsilon$ and $\delta$ values are presented in Figure \ref{fig:utility}. There are notable drops in TVD as $\epsilon$ increases from 0.1 to 1 and  from 1 to 10. For the Gaussian mechanisms of $(\epsilon,\delta)$-DP and $(\epsilon,\delta)$-pDP, the effect of $\delta$ on TVD is the most obvious when $\epsilon$ is around 1. Taken together with the DR-HA results in Figure \ref{fig:rhole}, $\epsilon$ around 1 seems to be a good choice to provide sufficient protection against DR-HA compared to no sanitization at all and acceptable utility when releasing low-dimensional marginals. In terms of the choice of $\delta$ in the cases of $(\epsilon,\delta)$-DP and $(\epsilon,\delta)$-pDP, the general recommendation of $O(n^{-1})$ is a good choice given its negligible impact on DR-HA and smaller impact than utility compared to $\epsilon$.

\vspace{-9pt} \section{Discussion}\label{sec:discussion}\vspace{-4pt} 
We derived the mathematical relationships between DR-HA in sanitized FDs via a randomized mechanism and  its privacy loss parameters. The relationships allow  practitioners to calculate DR-HA directly given a privacy loss without resorting to numerical evaluations (e.g., MC approaches). The  relationships also connect the rather abstract concept of privacy parameters to a privacy risk metric that is more concrete and intuitive, offering practitioners an additional perspective when choosing privacy loss parameters when sanitizing data. The R code used in the experiments are available at \url{https://github.com/zhao-xingyuan/DR-HA}. 

We focused on the Laplace and Gaussian mechanisms to demonstrate  the relationship between  DR-HA and  privacy loss. We plan to derive relationships between the DR-HA measures in Section \ref{sec:DR-HA} and privacy loss for other count sanitization mechanisms  with better utility than the Laplace and Gaussian mechanisms, such as the geometric mechanism. If sanitized counts are more accurate in one mechanism, it is expected that DR-HA would be higher as the sanitized data are closer to the original, despite the same privacy guarantees  across different mechanisms in the DP setting, but we expect the $S$ shape of the relationships to remain. 

We focused on private releasing of FDs, where the data are counts in nature. Adversaries learn about values of sensitive attributes, which may be non-numerical (such as HIV status), via HA on released FDs. Our methodology and results do not apply directly  to other types of data unless they can be converted to FDs. For example, continuous data can be discretized into bins/buckets, based on which FDs can be formed  and our methodology and results can then be applied, understanding that information loss would occur during the coarsening/discretization process.

We focused on DR-HA.  Future work may extend the results to other types of privacy risks due to HA or DR due to other types of attacks.  For example, while  DR-HA is zero in heterogeneous cells  per definition, it does not mean that DR does not exist from other attacks than HA. Adversaries may apply the plurality rule to predict sensitive information  (hard thresholding) or  be content with  probabilistic conclusions regarding sensitive information  (soft thresholding) of the records in a heterogeneous cell.   We illustrate, using the Adult data, the hard and soft thresholding approaches in assessing DR in heterogeneous cells and calculate the relationships between the combined DR from HA in homogeneous cells and using thresholding in heterogeneous cells after DP sanitization vs privacy loss parameters in the supplementary materials. The relations are also somewhat S-shaped but flatter compared to the relations in Section \ref{sec:experiment}, especially for soft-thresholding. The upper asymptotes are still 100\%, but the lower asymptotes shift to $>0.25$, as expected, due to the additional DR from the heterogeneous cells after sanitization.

\vspace{-12pt} 
\subsection*{Acknowledgments}\vspace{-4pt} 
\small We thank an associate editor and two referees for their comments and suggestions. We also thank Gaofei Zhang for initial discussions and work in this paper.

\normalsize
\vspace{-12pt}
\bibliographystyle{IEEEtranN}
\bibliography{reference}

% Generated by IEEEtranN.bst, version: 1.14 (2015/08/26)
\begin{thebibliography}{31}
\providecommand{\natexlab}[1]{#1}
\providecommand{\url}[1]{#1}
\csname url@samestyle\endcsname
\providecommand{\newblock}{\relax}
\providecommand{\bibinfo}[2]{#2}
\providecommand{\BIBentrySTDinterwordspacing}{\spaceskip=0pt\relax}
\providecommand{\BIBentryALTinterwordstretchfactor}{4}
\providecommand{\BIBentryALTinterwordspacing}{\spaceskip=\fontdimen2\font plus
\BIBentryALTinterwordstretchfactor\fontdimen3\font minus
  \fontdimen4\font\relax}
\providecommand{\BIBforeignlanguage}[2]{{%
\expandafter\ifx\csname l@#1\endcsname\relax
\typeout{** WARNING: IEEEtranN.bst: No hyphenation pattern has been}%
\typeout{** loaded for the language `#1'. Using the pattern for}%
\typeout{** the default language instead.}%
\else
\language=\csname l@#1\endcsname
\fi
#2}}
\providecommand{\BIBdecl}{\relax}
\BIBdecl

\bibitem[Duncan and Lambert(1989)]{duncan1989risk}
G.~Duncan and D.~Lambert, ``The risk of disclosure for microdata,''
  \emph{Journal of Business \& Economic Statistics}, vol.~7, no.~2, pp.
  207--217, 1989.

\bibitem[Skinner and Elliot(2002)]{skinner2002measure}
C.~J. Skinner and M.~Elliot, ``A measure of disclosure risk for microdata,''
  \emph{Journal of the Royal Statistical Society: series B (statistical
  methodology)}, vol.~64, no.~4, pp. 855--867, 2002.

\bibitem[Yancey et~al.(2002)Yancey, Winkler, and Creecy]{yancey2002disclosure}
W.~E. Yancey, W.~E. Winkler, and R.~H. Creecy, ``Disclosure risk assessment in
  perturbative microdata protection,'' in \emph{Inference control in
  statistical databases}.\hskip 1em plus 0.5em minus 0.4em\relax Springer,
  2002, pp. 135--152.

\bibitem[Reiter(2005)]{reiter2005estimating}
J.~P. Reiter, ``Estimating risks of identification disclosure in microdata,''
  \emph{Journal of the American Statistical Association}, vol. 100, no. 472,
  pp. 1103--1112, 2005.

\bibitem[Hundepool et~al.(2012)Hundepool, Domingo-Ferrer, Franconi, Giessing,
  Nordholt, Spicer, and De~Wolf]{hundepool2012statistical}
A.~Hundepool, J.~Domingo-Ferrer, L.~Franconi, S.~Giessing, E.~S. Nordholt,
  K.~Spicer, and P.-P. De~Wolf, \emph{Statistical disclosure control}.\hskip
  1em plus 0.5em minus 0.4em\relax John Wiley \& Sons, 2012.

\bibitem[Hu and Savitsky(2018)]{hu2018bayesian}
J.~Hu and T.~D. Savitsky, ``Bayesian data synthesis and disclosure risk
  quantification: An application to the consumer expenditure surveys,''
  \emph{arXiv:1809.10074}, 2018.

\bibitem[Dwork et~al.(2006{\natexlab{a}})Dwork, McSherry, Nissim, and
  Smith]{dwork2006calibrating}
C.~Dwork, F.~McSherry, K.~Nissim, and A.~Smith, ``Calibrating noise to
  sensitivity in private data analysis,'' in \emph{Theory of cryptography
  conference}.\hskip 1em plus 0.5em minus 0.4em\relax Springer, 2006, pp.
  265--284.

\bibitem[McSherry and Talwar(2007)]{mcsherry2007mechanism}
F.~McSherry and K.~Talwar, ``Mechanism design via differential privacy,'' in
  \emph{FOCS'07. 48th Annual IEEE Symposium on}.\hskip 1em plus 0.5em minus
  0.4em\relax IEEE, 2007, pp. 94--103.

\bibitem[Dwork et~al.(2010)Dwork, Rothblum, and Vadhan]{dwork2010boosting}
C.~Dwork, G.~N. Rothblum, and S.~Vadhan, ``Boosting and differential privacy,''
  in \emph{2010 IEEE 51st Annual Symposium on Foundations of Computer
  Science}.\hskip 1em plus 0.5em minus 0.4em\relax IEEE, 2010, pp. 51--60.

\bibitem[Kasiviswanathan et~al.(2011)Kasiviswanathan, Lee, Nissim,
  Raskhodnikova, and Smith]{kasiviswanathan2011can}
S.~P. Kasiviswanathan, H.~K. Lee, K.~Nissim, S.~Raskhodnikova, and A.~Smith,
  ``What can we learn privately?'' \emph{SIAM Journal on Computing}, vol.~40,
  no.~3, pp. 793--826, 2011.

\bibitem[Abadi et~al.(2016)Abadi, Chu, Goodfellow, McMahan, Mironov, Talwar,
  and Zhang]{abadi2016deep}
M.~Abadi, A.~Chu, I.~Goodfellow, H.~B. McMahan, I.~Mironov, K.~Talwar, and
  L.~Zhang, ``Deep learning with differential privacy,'' in \emph{Proceedings
  of the 2016 ACM SIGSAC Conference on Computer and Communications Security},
  2016, pp. 308--318.

\bibitem[Mironov(2017)]{mironov2017renyi}
I.~Mironov, ``R{\'e}nyi differential privacy,'' in \emph{2017 IEEE 30th
  Computer Security Foundations Symposium (CSF)}.\hskip 1em plus 0.5em minus
  0.4em\relax IEEE, 2017, pp. 263--275.

\bibitem[Dwork(2008)]{dwork2008differential}
C.~Dwork, ``Differential privacy: A survey of results,'' in \emph{International
  Conference on Theory and Applications of Models of Computation}.\hskip 1em
  plus 0.5em minus 0.4em\relax Springer, 2008, pp. 1--19.

\bibitem[Dwork et~al.(2019)Dwork, Kohli, and Mulligan]{dwork2019differential}
C.~Dwork, N.~Kohli, and D.~Mulligan, ``Differential privacy in practice: Expose
  your epsilons!'' \emph{Journal of Privacy and Confidentiality}, vol.~9,
  no.~2, 2019.

\bibitem[Lee and Clifton(2011)]{Clifton2011}
J.~Lee and C.~Clifton, ``How much is enough? choosing $\varepsilon$ for
  differential privacy,'' in \emph{International Conference on Information
  Security}.\hskip 1em plus 0.5em minus 0.4em\relax Springer, 2011, pp.
  325--340.

\bibitem[McClure and Reiter(2012)]{mcclure2012}
D.~McClure and J.~P. Reiter, ``Differential privacy and statistical disclosure
  risk measures: An investigation with binary synthetic data.'' \emph{Trans.
  Data Priv.}, vol.~5, no.~3, pp. 535--552, 2012.

\bibitem[Hsu et~al.(2014)Hsu, Gaboardi, Haeberlen, Khanna, Narayan, Pierce, and
  Roth]{roth2014}
J.~Hsu, M.~Gaboardi, A.~Haeberlen, S.~Khanna, A.~Narayan, B.~C. Pierce, and
  A.~Roth, ``Differential privacy: An economic method for choosing epsilon,''
  in \emph{2014 IEEE 27th Computer Security Foundations Symposium}, 2014, pp.
  398--410.

\bibitem[Abowd and Schmutte(2015)]{abowd2015}
J.~M. Abowd and I.~M. Schmutte, ``Revisiting the economics of privacy:
  Population statistics and confidentiality protection as public goods,''
  \url{https://ecommons.cornell.edu/handle/1813/39081}, 2015.

\bibitem[Nissim et~al.(2017)Nissim, Steinke, Wood, Altman, Bembenek, Bun,
  Gaboardi, O’Brien, and Vadhan]{nissim2017differential}
K.~Nissim, T.~Steinke, A.~Wood, M.~Altman, A.~Bembenek, M.~Bun, M.~Gaboardi,
  D.~R. O’Brien, and S.~Vadhan, ``Differential privacy: A primer for a
  non-technical audience,'' in \emph{Privacy Law Scholars Conf}, vol.~3, 2017.

\bibitem[Dwork et~al.(2017)Dwork, Smith, Steinke, and Ullman]{dwork2017exposed}
C.~Dwork, A.~Smith, T.~Steinke, and J.~Ullman, ``Exposed! a survey of attacks
  on private data,'' \emph{Ann. Rev. of Stats. and Its Appl.}, vol.~4, pp.
  61--84, 2017.

\bibitem[Holohan et~al.(2017)Holohan, Antonatos, Braghin, and
  Mac~Aonghusa]{holohan2017k}
N.~Holohan, S.~Antonatos, S.~Braghin, and P.~Mac~Aonghusa, ``($k$, $\epsilon
  $)-anonymity: $ k $-anonymity with $\epsilon$-differential privacy,''
  \emph{arXiv:1710.01615}, 2017.

\bibitem[Chen et~al.(2017{\natexlab{a}})Chen, Chen, Tsou, Yu, Tai, Li, Huang,
  and Lin]{chen2017evaluating}
H.-L. Chen, J.-Y. Chen, Y.-T. Tsou, C.-M. Yu, B.-C. Tai, S.-C. Li, Y.~Huang,
  and C.-M. Lin, ``Evaluating the risk of data disclosure using noise
  estimation for differential privacy,'' in \emph{2017 IEEE 22nd Pacific Rim
  International Symposium on Dependable Computing (PRDC)}.\hskip 1em plus 0.5em
  minus 0.4em\relax IEEE, 2017, pp. 339--347.

\bibitem[Chen et~al.(2017{\natexlab{b}})Chen, Yu, Tai, Li, Tsou, Huang, and
  Lin]{chen2017data}
K.-C. Chen, C.-M. Yu, B.-C. Tai, S.-C. Li, Y.-T. Tsou, Y.~Huang, and C.-M. Lin,
  ``Data-driven approach for evaluating risk of disclosure and utility in
  differentially private data release,'' in \emph{2017 IEEE 31st International
  Conference on Advanced Information Networking and Applications (AINA)}.\hskip
  1em plus 0.5em minus 0.4em\relax IEEE, 2017, pp. 1130--1137.

\bibitem[Dwork et~al.(2006{\natexlab{b}})Dwork, Kenthapadi, McSherry, Mironov,
  and Naor]{dwork2006our}
C.~Dwork, K.~Kenthapadi, F.~McSherry, I.~Mironov, and M.~Naor, ``Our data,
  ourselves: Privacy via distributed noise generation,'' in \emph{Annual
  international conference on the theory and applications of cryptographic
  techniques}.\hskip 1em plus 0.5em minus 0.4em\relax Springer, 2006, pp.
  486--503.

\bibitem[Machanavajjhala et~al.(2008)Machanavajjhala, Kifer, Abowd, Gehrke, and
  Vilhuber]{machanavajjhala2008privacy}
A.~Machanavajjhala, D.~Kifer, J.~Abowd, J.~Gehrke, and L.~Vilhuber, ``Privacy:
  Theory meets practice on the map,'' in \emph{Proceedings of the 2008 IEEE
  24th International Conference on Data Engineering}.\hskip 1em plus 0.5em
  minus 0.4em\relax IEEE Computer Society, 2008, pp. 277--286.

\bibitem[Liu(2019)]{liu2018generalized}
F.~Liu, ``Generalized gaussian mechanism for differential privacy,'' \emph{IEEE
  Transactions on Knowledge and Data Engineering}, vol. 31 (4), pp. 747 --756,
  2019.

\bibitem[Dwork et~al.(2014)Dwork, Roth, et~al.]{dwork2014algorithmic}
C.~Dwork, A.~Roth \emph{et~al.}, ``The algorithmic foundations of differential
  privacy,'' \emph{Foundations and Trends in Theoretical Computer Science},
  vol.~9, no. 3--4, pp. 211--407, 2014.

\bibitem[Dalenius(1986)]{dalenius1986finding}
T.~Dalenius, ``Finding a needle in a haystack or identifying anonymous census
  records,'' \emph{Journal of official statistics}, vol.~2, no.~3, p. 329,
  1986.

\bibitem[Ghosh et~al.(2009)Ghosh, Roughgarden, and
  Sundararajan]{ghosh2009universally}
A.~Ghosh, T.~Roughgarden, and M.~Sundararajan, ``Universally utility-maximizing
  privacy mechanisms,'' in \emph{Proceedings of the forty-first annual ACM
  symposium on Theory of computing}, 2009, pp. 351--360.

\bibitem[Kohavi and Becker(1996)]{Adult1996}
R.~Kohavi and B.~Becker, ``Qualitative bankruptcy data set,''
  \url{https://archive.ics.uci.edu/ml/datasets/Adult}, 1996.

\bibitem[Martin et~al.(2014)Martin, Uthayakumar, and
  M.Nadarajan]{Bankruptcy2014}
A.~Martin, J.~Uthayakumar, and M.Nadarajan, ``Adult data set,''
  \url{https://archive.ics.uci.edu/ml/datasets/qualitative_bankruptcy}, 2014.

\end{thebibliography}

\end{document}

\setcounter{page}{1}
\setcounter{figure}{0}
\setcounter{table}{0}
\setcounter{equation}{0}
\setcounter{section}{0}

\renewcommand{\thepage}{S\arabic{page}}
\renewcommand{\thesection}{S\arabic{section}}
\renewcommand{\thetable}{S\arabic{table}}
\renewcommand{\thefigure}{S\arabic{figure}}
\renewcommand{\theequation}{S\arabic{equation}}
\renewcommand{\thecor}{S\arabic{thm}}

\onecolumn

\begin{center} \LARGE\bf{Supplementary Materials to \\
``Disclosure Risk from Homogeneity Attack in Differentially Privately Sanitized Frequency Distribution''} \\
\vspace{12pt}
\large{\textbf{ Fang Liu,\footnote{\noindent Fang Liu (email: fang.liu.131@nd.edu) is Professor and Xingyuan Zhao is a doctoral student in the Department of Applied and Computational Mathematics and Statistics at the University of Notre Dame. This work was supported by NSF Award \#1717417.} Xingyuan Zhao}\\
\vspace{6pt}
\normalsize{Applied and Computational Mathematics and Statistics}\\
\vspace{3pt}
\normalsize{University of Notre Dame, Notre Dame, IN 46556}}
\end{center}

\normalsize

\normalsize
\section{\large  Proofs of Theorem 1 and Corollary 2}\label{rho.lap.proof}
\begin{proof}
We first examine the case of $K=2$, and then extend to the case of $K\ge2$. 
Per Definition 6, the homogeneity attack occurs in Scenarios 1 and 8. In Scenario 1, the local DR-HA on sensitive attribute $Y$ in cell $\C_i$ is
\begin{align*}
\rho^{\text{l}}_i&=(\Pr(n_{i,1}\!=\!n_i)\!+\!\Pr(n_{i,1}\!=\!0))\Pr(n_{i,0}+e_{i,0}<0.5|n_{i,1}\!=\!n_i)\Pr(n_{i,1}+e_{i,1}\geqslant0.5|n_{i,1}\!=\!n_i)\notag\\
&=(\Pr(n_{i,1}=n_{i})+\Pr(n_{i,1}\!=\!0))\Pr(e_{i,0}<0.5)\Pr(e_{i,1}\geqslant0.5-n_i)\notag\\
&=\left(p_i^{n_i}+(1-p_i)^{n_i}\right)\left(1-0.5\exp(-0.5\epsilon)\right)\left(1-0.5\exp((0.5-n_i)\epsilon)\right).
\end{align*}
WLOS, we examine $(n_{i,0},n_{i,1})=(1,n_i-1)$ for Scenario 8, where
\small\begin{align*}
&\rho^{\text{l}}_i\!<\!\mathbbm{1}(n_i\!\ge\!2)\Pr(n_{i,0},n_{i,1})\!\neq\!(0,n_i)\big\{\!\Pr(n_{i,1}\!+\!e_{i,1}\!\geq\!0.5|n_{i,1}\!=\!n_i\!-\!1)\Pr(n_{i,0}\!+\!e_{i,0}\!<\!0.5|n_{i,0}\!=\!1)\\
&+\Pr(n_{i,1}\!+\!e_{i,1}\!<\!0.5|n_{i,1}\!=\!n_i\!-\!1)\Pr(n_{i,0}\!+\!e_{i,0}\!\geq\!0.5|n_{i,0}\!=\!1)\big\}\\
=\;&\!\mathbbm{1}(n_i\!\ge\!2)(1\!-\!p_i^{n_i}\!-\!(1\!-\!p_i)^{n_i})\!\left(\Pr(e_{i,1}\!\!\geq\!\!1.5\!-\!n_i)\Pr(e_{i,0}\!\!<\!\!-0.5)\!+\!\Pr(e_{i,1}\!\!<\!\!1.5\!-\!n_i)\Pr(e_{i,0}\!\!\geq\!\!-0.5)\right)\\
=\;&\mathbbm{1}(n_i\!\ge\!2)\left(1\!-\!p_i^{n_i}\!-\!(1\!-\!p_i)^{n_i}\right)[\left(\!1\!-\!0.5\exp(\epsilon(1.5\!-\!n_i))\right)\left(0.5\exp(-0.5\epsilon)\right)+0.5\exp((1.5\!-\!n_i)\epsilon)(1\!-\!0.5\exp(-0.5\epsilon))].
\end{align*}
\normalsize The expected DR-HA of $Y$ in cell $\C_i$ in the sanitized data is thus given by
\begin{align*}
    \rho^{\text{e}}_i<&\left(p_i^{n_i}+(1-p_i)^{n_i}\right)\left(1-0.5\exp(-0.5\epsilon)\right)\left(1-0.5\exp((0.5-n_i)\epsilon)\right)+\\
    &\mathbbm{1}(n_i\!\ge\!2)\left(1-p_i^{n_i}-(1-p_i)^{n_i}\right)[\left(1-0.5\exp(\epsilon(1.5-n_i))\right)\left(0.5\exp(-0.5\epsilon)\right)+\\
&0.5\exp((1.5\!-\!n_i)\epsilon)(1\!-\!0.5\exp(-0.5\epsilon))].
\end{align*}
Assuming $p_i\sim\mbox{beta}(\alpha_1,\alpha_2)$, the shrinkage disclosure risk in cell $\C_i$ is
\begin{align*}
\rho^{\text{s}}_i\!<&\left(1\!-\!\frac{1}{2}e^{-0.5\epsilon}\!\right)\!\!\left(\!1\!-\!\frac{1}{2}e^{(0.5-n_i)\epsilon}\!\right)\int_0^1\!(p_i^{n_i}\!+\!(1\!-\!p_i)^{n_i})\frac{p_i^{\alpha_1-1}(1\!-\!p_i)^{\alpha_2-1}}{B(\alpha_1,\alpha_2)}dp_i+\!\mathbbm{1}(n_i\!\ge\!2)\\
&\times\![(1\!-\!0.5\exp(\epsilon(1.5-n_i)))(0.5\exp(-0.5\epsilon))\!+\!0.5\exp((1.5\!-\!n_i)\epsilon)(1\!-\!0.5\exp(-0.5\epsilon))]\\
&\times\!\int_0^1\!\!\!\left(1\!-\!p_i^{n_i}\!-\!(1\!-\!p_i)^{n_i}\right)\!\frac{p_i^{\alpha_1-1}(1\!-\!p_i)^{\alpha_2-1}}{B(\alpha_1,\alpha_2)}dp_i.\\
=&\frac{(1-\frac{1}{2}e^{-\frac{\epsilon}{2}})(1\!-\!\frac{1}{2}e^{(0.5-n_i)\epsilon})}{B(\alpha_1,\alpha_2)}\!\!\int_{0}^{1}(p_i^{n_i+\alpha_1-1}(1\!-\!p_i)^{\alpha_2-1}\!+\!p_i^{\alpha_1-1}(1\!-\!p_i)^{n_i+\alpha_2-1})dp_i+\\
&\mathbbm{1}(n_i\!\ge\!2)\left(1-\!\frac{1}{B(\alpha_1,\alpha_2)}\int_{0}^{1}(p_i^{n_i+\alpha_1-1}(1\!-\!p_i)^{\alpha_2-1}\!+\!p_i^{\alpha_1-1}(1\!-\!p_i)^{n_i+\alpha_2-1})dp_i\right)\times\\
&[\left(1-0.5\exp(\epsilon(1.5-n_i))\right)\left(0.5\exp(-0.5\epsilon)\right)+0.5\exp((1.5\!-\!n_i)\epsilon)(1\!-\!0.5\exp(-0.5\epsilon))].\\
=&\frac{(1\!-\!\frac{1}{2}e^{-\frac{\epsilon}{2}})\left(\!1\!-\!\frac{1}{2}e^{(0.5-n_i)\epsilon}\right)}{B(\alpha_1,\alpha_2)}(B(n_i\!+\!\alpha_1,\alpha_2)\!+\!B(\alpha_1,n_i\!+\!\alpha_2))+\mathbbm{1}(n_i\!\ge\!2)\left(1-\!\frac{B(n_i\!+\!\alpha_1,\alpha_2)\!+\!B(\alpha_1,n_i\!+\!\alpha_2)}{B(\alpha_1,\alpha_2)}\!\right)\\
&\times[\left(1-0.5\exp(\epsilon(1.5-n_i))\right)\left(0.5\exp(-0.5\epsilon)\right)+0.5\exp((1.5\!-\!n_i)\epsilon)(1\!-\!0.5\exp(-0.5\epsilon))].
\end{align*}

When $K\ge2$, the DR-HA of cell $\C_i$ after sanitization in Scenario 1 is
\begin{align*}
\rho^{\text{l}}_i&=\textstyle\sum_{k=1}^{K}\!\!\left(\Pr(n_{i,k}\!=\!n_i)\Pr(n_{i,k}\!+\!e_{i,k}\!\geqslant\!0.5|n_{i,k}\!=\!n_i)\!\prod_{k'\neq k}\Pr(n_{i,k'}\!+\!e_{i,k'}\!<\!0.5|n_{i,k}\!=\!n_i)\right)\\
&=\textstyle\sum_{k=1}^{K}\left(p_{i,k}^{n_i}\Pr(e_{i,k}\geqslant0.5-n_i)
\prod_{k'\neq k}\Pr(e_{i,k'}<0.5)\right)\\
&=\textstyle\sum_{k=1}^{K}p_{i,k}^{n_i}\left(1-0.5\exp(-0.5\epsilon)\right)^{K-1}\left(1-0.5\exp((0.5-n_i)\epsilon)\right);
\end{align*}
WLOS, let $\{n_{i,1},\dots,n_{i,K}\}=\{n_i-1,1,0,\dots,0\}$ in Scenario 8. Then\small
\begin{align*}
&\rho^{\text{l}}_i<\;\textstyle\!\mathbbm{1}(n_i\!\ge\!2)\!\left(1\!-\!\sum_{k=1}^{K}\!p_{i,k}^{n_i}\!\right)\![\Pr(n_{i,1}\!+\!e_{i,1}\!\geq\!0.5|n_{i,1}\!=\!n_i\!-\!1)\Pr(n_{i,2}\!+\!e_{i,2}\!<\!0.5|n_{i,2}\!=\!1)\times\\
&\prod_{t=3}^K\Pr(n_{i,t}\!+\!e_{i,t}\!<\!0.5|n_{i,t}\!=\!0)\!+\!\Pr(n_{i,1}\!\!+\!\!e_{i,1}\!<\!0.5|n_{i,1}\!=\!n_i\!-\!1)\Pr(n_{i,2}\!+\!e_{i,2}\!\geq\!0.5|n_{i,2}\!=\!1)\prod_{t=3}^K\Pr(n_{i,t}\!+\!e_{i,t}\!<\!0.5|n_{i,t}\!=\!0)\\
\textstyle=&\mathbbm{1}(n_i\!\ge\!2)\!\left(1\!-\!\sum_{k=1}^{K}\!p_{i,k}^{n_i}\right)\![\Pr(e_{i,1}\geq 1.5-n_i)\Pr(e_{i,2}<-0.5)\prod_{t=3}^K\Pr(e_{i,t}<0.5)\!+\!\Pr(e_{i,1}< 1.5-n_i)\Pr(e_{i,2}\geq-0.5)\prod_{t=3}^K\Pr(e_{i,t}<0.5)]\\
=&\mathbbm{1}(n_i\!\ge\!2)\!\left(\!1\!-\!\sum_{k=1}^{K}\!p_{i,k}^{n_i}\!\!\right)\!\!\bigg\{\left(\!1\!-\!\frac{1}{2}\exp((1.5\!-\!n_i)\epsilon)\!\right)\!\!\left(\!\frac{1}{2}\exp(-0.5\epsilon)\right)\!\!\left(\!1\!-\!\frac{1}{2}\exp(-0.5\epsilon)\!\right)^{\!K-2}\!\!\!\!+\!\left(\!\frac{1}{2}\exp((1.5\!-\!n_i)\epsilon)\right)\!\!\left(\!1\!-\!\frac{1}{2}\exp(-0.5\epsilon)\!\right)^{\!K\!-\!1}\bigg\}\\
=&\mathbbm{1}(n_i\!\ge\!2)\!\left(\!1\!-\!\sum_{k=1}^{K}p_{i,k}^{n_i}\!\right)\!(1\!-\!0.5e^{-0.5\epsilon})^{K-2}\times0.5(e^{-0.5\epsilon}+e^{(1.5-n_i)\epsilon}-e^{(1-n_i)\epsilon}).
\end{align*}
\normalsize The expected DR-HA of $Y$ in cell $\C_i$ in the sanitized data is thus
\begin{align*}
\rho^{\text{e}}_i<&\!\bigg(\!\sum_{k=1}^{K}p_{i,k}^{n_i}\!\bigg)\big(1\!-\!0.5e^{-0.5\epsilon}\big)^{K\!-\!1}\!(1\!-\!0.5e^{(0.5-n_i)\epsilon})\!+\!\bigg\{\! \mathbbm{1}(n_i\!\ge\!2)\!\left(\!1\!-\!\sum_{k=1}^{K}p_{i,k}^{n_i}\!\right)\!(1\!-\!0.5e^{-0.5\epsilon})^{K\!-\!2}\\
&\times0.5(e^{-0.5\epsilon}+e^{(1.5-n_i)\epsilon}-e^{(1-n_i)\epsilon})\bigg\}.
\end{align*}
Assuming $(p_{i,1},\ldots,p_{i,K})\overset{\text{ind}}{\sim}\mbox{Dirichlet}(\alpha_{1},\ldots,\alpha_{K})$, the shrinkage DR-HA in cell $\C_i$ is
\small
\begin{align*}
\rho^{\text{s}}_i\!<&\left(1\!-\!0.5e^{-0.5\epsilon}\right)^{K-1}\!\!\left(\!1\!-\!0.5e^{(0.5-n_i)\epsilon}\right)
\int\!\sum_{k=1}^{K}p_{i,k}^{n_i}\!\frac{\Gamma(\sum_{k=1}^K \alpha_{k})}{\prod_{k=1}^K\Gamma(\alpha_{k})}\!\prod_{k=1}^{K}p_{i,k}^{\alpha_k-1}d\mathbf{p}+\!\\
&\mathbbm{1}(n_i\!\ge\!2)(1\!-\!0.5e^{-0.5\epsilon})^{K\!-\!2}\times0.5(e^{-0.5\epsilon}\!+\!e^{(1.5-n_i)\epsilon}\!-\!e^{(1-n_i)\epsilon})\int\!\left(\!1\!-\!\sum_{k=1}^{K}p_{i,k}^{n_i}\!\right)\!\frac{\Gamma(\sum_{k=1}^K \alpha_{k})}{\prod_{k=1}^K\!\Gamma(\alpha_{k})}\prod_{k=1}^{K}p_{i,k}^{\alpha_k-1}d\mathbf{p}\\
=&\left[
\sum_{k=1}^{K}\!\left(\!\int\!p_{i,k}^{n_i}\!\!\prod_{k=1}^{K}p_{i,k}^{\alpha_k-1}\!\frac{\Gamma(\sum_{k}\alpha_{k}\!+\!n_i)}{\Gamma(\alpha_k\!+\!n_i)\!\prod_{k'\ne k}\! \Gamma(\alpha_{k'})}d\mathbf{p}\!\!\right) \!\frac{\Gamma(\sum_{k}\alpha_{k})\Gamma(\alpha_k\!+\!n_i)}{\Gamma(\sum_{k}\alpha_{k}+n_i)\Gamma(\alpha_k)}\!\right]\times\\
&\left(1\!-\!e^{-0.5\epsilon}/2\right)^{\!K-1}\!\left(1\!-\!e^{(0.5-n_i)\epsilon}/2\right)+\mathbbm{1}(n_i\!\ge\!2)\!(1\!-\!0.5e^{-0.5\epsilon})^{\!K\!-\!2}\times\\
&0.5(e^{-0.5\epsilon}\!+\!e^{(1.5-n_i)\epsilon}\!-\!e^{(1-n_i)\epsilon})\!\left(\!1-\!\int\!\!\left(\!\sum_{k=1}^{K}p_{i,k}^{n_i}\!\!\right)\!\!\frac{\Gamma(\sum_{k=1}^K \alpha_{k})}{\prod_{k=1}^K\Gamma(\alpha_{k})}\prod_{k=1}^{K}p_{i,k}^{\alpha_k-1}d\mathbf{p}\!\right)\\
=&\left(\sum_{k=1}^{K}\frac{\Gamma(\sum_k\alpha_k)\Gamma(\alpha_k+n_i)}{\Gamma(\sum_k\alpha_k+n_i)\Gamma(\alpha_k)}\right)\left(1-0.5e^{-0.5\epsilon}\right)^{K-1}\left(1-0.5e^{(0.5-n_i)\epsilon}\right)+\mathbbm{1}(n_i\!\ge\!2)\times\\
&(1\!-\!0.5e^{-0.5\epsilon})^{\!K\!-\!2}\times0.5(e^{-0.5\epsilon}\!+\!e^{(1.5-n_i)\epsilon}\!-\!e^{(1-n_i)\epsilon})\!\left(\!1\!-\!\sum_{k=1}^{K}\frac{\Gamma(\sum_k\alpha_k)\Gamma(\alpha_k\!+\!n_i)}{\Gamma(\sum_k\alpha_k\!+\!n_i)\Gamma(\alpha_k)}\right).
\end{align*}
\end{proof}

\vspace{-24pt}
\section{\large  EB estimation of hyperparameters in Dirichlet Prior} \label{MoM.proof}\vspace{-3pt}
We examine the case of $K=2$ first and then generalize it to the general case of $K\ge2$. When $K=2$,
\begin{align*}
n_{i,1}|p_{i,1}\sim\mbox{Binomial}(n_i,p_{i,1})\\
p_{i,1}\sim\mbox{Beta}(\alpha_1,\alpha_2)\mbox{ for }i=1,\dots,M.
\end{align*}
For each cell $\C_i$, 
$$f(n_{i,1}|\alpha_1,\alpha_2)=\binom{n_i}{n_{i,1}}\frac{B(n_{i,1}+\alpha_1,n_i-n_{i,1}+\alpha_2)}{B(\alpha_1,\alpha_2)},$$
from which we can get
$$E(n_{i,1})=\frac{n_i\alpha_1}{\alpha_1+\alpha_2}\mbox{ and }\mbox{Var}(n_{i,1})=\frac{n_i\alpha_1\alpha_2(\alpha_1+\alpha_2+n_i)}{(\alpha_1+\alpha_2)^2(\alpha_1+\alpha_2+1)}.$$
The variance can be rewritten as 
\begin{align*}
\mbox{Var}(n_{i,1})
=\frac{n_i\alpha_1\alpha_2}{(\alpha_1+\alpha_2)(\alpha_1+\alpha_2+1)}\!+\!\frac{n_i^2\alpha_1\alpha_2}{(\alpha_1+\alpha_2)^2(\alpha_1+\alpha_2+1)}.
\end{align*}
Let $\sum_{i=1}^N n_{i,1}=n_1$, $\sum_{i=1}^N n_i=n$. Assuming $n_{i,1}$ are independent, we have
$$E(n_1)=\frac{n\alpha_1}{\alpha_1+\alpha_2}\mbox{ and }\mbox{Var}(n_1)=\frac{n\alpha_1\alpha_2}{(\alpha_1+\alpha_2)(\alpha_1+\alpha_2+1)}\!+\!\frac{\sum n_i^2\alpha_1\alpha_2}{(\alpha_1+\alpha_2)^2(\alpha_1+\alpha_2+1)}.$$
Let $E(n_1)$ and $\mbox{Var}(n_1)$ equal to their empirical moments $n\hat{p}$ and $s^2=n\hat{p}(1-\hat{p})$, respectively,  where $\hat{p}=n_1/n$, we have
$$\hat{\alpha}_1/(\hat{\alpha}_1+\hat{\alpha}_2)= \hat{p}$$
$$\big(n\hat{p}\hat{\alpha}_2+\hat{p}(1-\hat{p})\sum n_i^2\big)/(\hat{\alpha}_1+\hat{\alpha}_2+1)=s^2,$$
from which we can solve for the MoM estimates of $\alpha_1$ and $\alpha_2$
$$\hat{\alpha}_1=\frac{\hat{p}^2(1-\hat{p})\sum n_i^2-\hat{p}s^2}{s^2-n\hat{p}(1-\hat{p})};\;\hat{\alpha}_2=\frac{\hat{p}(1-\hat{p})^2\sum n_i^2-(1-\hat{p})s^2}{s^2-n\hat{p}(1-\hat{p})}.$$

When $K\geq2$,
\begin{align*}
(n_{i,1},\dots,n_{i,K})|p_{i,1},\dots,p_{i,K}\sim\mbox{multinomial}(n_i,(p_{i,1},\dots,p_{i,K}))\\
(p_{i,1},\dots,p_{i,K})\sim\mbox{Dirichlet}(\alpha_1,\dots,\alpha_K)\mbox{ for }i=1,\dots,M
\end{align*}
Let $\alpha.=\sum_{k=1}^K\alpha_k$ and $p_k=\alpha_k/\alpha.$. For each cell $\C_i$, 
\begin{align*}
&f(n_{i,1},\dots,n_{i,K}|\alpha_1,\dots,\alpha_K)=\frac{\Gamma(\alpha.)}{\Gamma(n_i+\alpha.)}\prod_{k=1}^K\frac{\Gamma(n_{i,k}+\alpha_k)}{\Gamma(\alpha_k)}\\
&E(n_{i,k})=n_ip_k\\
&\mbox{Var}(n_{i,k})=n_ip_k(1-p_k)\left(\frac{n_i+\alpha.}{1+\alpha.}\right);\;
\mbox{Cov}(n_{i,k},n_{i,j})=-n_ip_kp_j\left(\frac{n_i+\alpha.}{1+\alpha.}\right),k\neq j,
\end{align*}
Let $\sum_{i=1}^N n_{i,k}=n_{\cdot k}$, $\sum_{i=1}^N n_i=n$ and assume $n_{i,k}$ is independent across $i$; we have
\begin{align*}
E(n_{\cdot k})=np_k, 
\mbox{Var}(n_{\cdot k})=p_k(1-p_k)\frac{\sum_in_i^2+\alpha.n}{1+\alpha.},\\
\mbox{Cov}(n_{\cdot k},n_{\cdot j})=\sum_i\mbox{Cov}(n_{i,k},n_{i,j})=-\frac{\alpha_k\alpha_j}{\alpha.^2}\left(\frac{\sum n_i^2+\alpha. n}{1+\alpha.}\right),k\neq j.
\end{align*}
Let $\Sigma$ be the covariance matrix of the frequencies $n_{.k}$; its elements can be estimated by
$\hat{\Sigma}_{kk}=(M-1)^{-1}\sum_{i=1}^N(n_{i,k}-\bar{n}_{\cdot k})^2$ and $\hat{\Sigma}_{kj}=(M-1)^{-1}\sum_{i=1}^N(n_{i,k}-\bar{n}_{\cdot k})(n_{i,j'}-\bar{n}_{\cdot j})$ for $k\ne j$,
where $\bar{n}_{\cdot k}=\sum_{i=1}^N n_{i,k}/M$. Set $E(n_{\cdot k})$ and $\mbox{Var}(n_{\cdot k})$ at their empirical moments $n_{\cdot k}$ and  $\hat{\Sigma}_{kk}$, respectively, we have $\hat{p}_k=\hat{\alpha}_k/\hat{\alpha}.= n_{\cdot k}/n$ and 
$\hat{p}_k\left(1-\hat{\theta}_k\right)(\sum n_i^2+\hat{\alpha}. n)/(1+\hat{\alpha}.)=\hat{\Sigma}_{kk}$, where $\hat{\alpha}.=\hat{\alpha}_1+\cdots+\hat{\alpha}_K$, from which we can obtain the MoM estimates of the hyperparameters $(\alpha_1,\dots,\alpha_K)$ 
$$\hat{\alpha}_k=\frac{\hat{p}_k^2(1-\hat{p}_k)\sum n_i^2-\hat{p}_k\hat{\Sigma}_{kk}}{\hat{\Sigma}_{kk}-\hat{p}_k(1-\hat{\theta}_k)n}.$$

\section{\large  Proofs of Theorem 3 and Corollary 4} \label{rho.gau.proof}
\begin{proof}
Similar to the proof in Section \ref{rho.lap.proof}, the DR-HA in Scenario 1
is 
$$\rho^{\text{l}}_i=[\Pr(n_{i,1}=n_i)+\Pr(n_{i,1}=0)]\Pr(e_{i,0}<0.5)\Pr(e_{i,1}\ge0.5-n_i),$$ 
where $e_{i,0}$ and $e_{i,1}$ follow $\mathcal{N}(0,\sigma^2)$. Plugging in  $\frac{1}{2}(1+\mbox{erf}(\frac{x}{\sigma\sqrt{2}}))$, the CDF of $\mathcal{N}(0,\sigma^2)$, we obtain the DR-HA in cell $\C_i$ $$\rho^{\text{l}}_i=\frac{1}{4}\left(p_i^{n_i}+(1-p_i)^{n_i}\right)\left(1+\mbox{erf}\left(\frac{1}{2\sqrt{2}\sigma}\right)\right)\left(1+\mbox{erf}\left(\frac{n_i-0.5}{\sqrt{2}\sigma}\right)\right).$$
In Scenario 8,
\begin{align*}
\rho^{\text{l}}_i<&\mathbbm{1}(n_i\!\ge\!2)(\Pr(\{n_{i,1}, n_{i,0}\}\!\neq\!\{n_i,0\})(\Pr(e_{i,1}\!\geq\!1.5\!-\!n_i)\Pr(e_{i,0}\!<\!-0.5)+\Pr(e_{i,1}\!<\!1.5\!-\!n_i)\Pr(e_{i,0}\!>\!-0.5))\\
=&\mathbbm{1}(n_i\!\ge\!2)\frac{1}{4}\left(1\!-\!p_i^{n_i}\!-\!(1\!-\!p_i)^{n_i}\!\right)\!\bigg\{\!\left(1\!-\!\mbox{erf}\!\left(\!\frac{1.5\!-\!n_i}{\sqrt{2}\sigma}\!\right)\!\right)\!\left(\!1\!+\!\mbox{erf}\!\left(\!\frac{-0.5}{\sqrt{2}\sigma}\!\right)\!\right)\!+\!\left(\!1\!+\!\mbox{erf}\left(\frac{1.5\!-\!n_i}{\sqrt{2}\sigma}\right)\!\right)\!\left(\!1\!+\!\mbox{erf}\!\left(\frac{0.5}{\sqrt{2}\sigma}\!\right)\!\right)\!\bigg\}.
\end{align*}
The expected DR-HA in cell $\C_i$ is thus
\begin{align*}
\rho^{\text{e}}_i&<\frac{1}{4}\left(p_i^{n_i}+(1-p_i)^{n_i}\right)\left(1+\mbox{erf}\left(\frac{1}{2\sqrt{2}\sigma}\right)\right)\left(1+\mbox{erf}\left(\frac{n_i-0.5}{\sqrt{2}\sigma}\right)\right)+\\
&\frac{1}{4}\mathbbm{1}(n_i\!\ge\!2)\left(1\!-\!p_i^{n_i}\!-\!(1\!-\!p_i)^{n_i}\!\right)\!\bigg\{\!\left(1\!-\!\mbox{erf}\!\left(\!\frac{1.5\!-\!n_i}{\sqrt{2}\sigma}\!\right)\!\right)\!\left(\!1\!+\!\mbox{erf}\!\left(\!\frac{-0.5}{\sqrt{2}\sigma}\!\right)\!\right)\!+\!\left(\!1\!+\!\mbox{erf}\left(\frac{1.5\!-\!n_i}{\sqrt{2}\sigma}\right)\!\right)\!\left(\!1\!+\!\mbox{erf}\!\left(\frac{0.5}{\sqrt{2}\sigma}\!\right)\!\right)\!\bigg\}.
\end{align*}
The shrinkage DR-HA in cell $\C_i$, assuming $p_i\sim\mbox{beta}(\alpha_1,\alpha_2)$, is 
\small \begin{align*}
\rho^{\text{s}}_i<&\frac{1}{4}\!\left(\!1\!+\!\mbox{erf}\left(\frac{1}{2\sqrt{2}\sigma}\right)\right)\!\left(\!1\!+\!\mbox{erf}\left(\frac{n_i-0.5}{\sqrt{2}\sigma}\right)\right)\int_{0}^{1}\!\!\left(p_i^{n_i}+(1-p_i)^{n_i}\right)\!\frac{p_i^{\alpha_1-1}(1-p_i)^{\alpha_2-1}}{B(\alpha_1,\alpha_2)}dp_i+\\
&\frac{1}{4}\mathbbm{1}(n_i\!\ge\!2)\!\bigg\{\!\!\left(1\!-\!\mbox{erf}\!\left(\!\frac{1.5\!-\!n_i}{\sqrt{2}\sigma}\!\right)\!\right)\!\left(\!1\!+\!\mbox{erf}\!\left(\!\frac{-0.5}{\sqrt{2}\sigma}\!\right)\!\right)\!+\!\left(\!1\!+\!\mbox{erf}\left(\frac{1.5\!-\!n_i}{\sqrt{2}\sigma}\right)\!\right)\!\left(\!1\!+\!\mbox{erf}\!\left(\frac{0.5}{\sqrt{2}\sigma}\!\right)\!\right)\!\!\bigg\}\\
&\times\int_0^1\!\!\left(1-p_i^{n_i}-(1-p_i)^{n_i}\right)\frac{p_i^{\alpha_1\!-\!1}(1\!-\!p_i)^{\alpha_2\!-\!1}}{B(\alpha_1,\alpha_2)}dp_i\!\\
=&\frac{1}{4B(\alpha_1,\alpha_2)}\left(B(n_i+\alpha_1,\alpha_2)+B(\alpha_1,n_i+\alpha_2)\right)\!\left(1\!+\!\mbox{erf}\left(\frac{1}{2\sqrt{2}\sigma}\right)\right)\!\left(1\!+\!\mbox{erf}\left(\frac{n_i\!-\!0.5}{\sqrt{2}\sigma}\right)\right)+\\
&\frac{1}{4}\mathbbm{1}(n_i\!\ge\!2)\!\bigg\{\!\!\left(1\!-\!\mbox{erf}\!\left(\!\frac{1.5\!-\!n_i}{\sqrt{2}\sigma}\!\right)\!\right)\!\left(\!1\!+\!\mbox{erf}\!\left(\!\frac{-0.5}{\sqrt{2}\sigma}\!\right)\!\right)\!+\!\left(\!1\!+\!\mbox{erf}\left(\frac{1.5\!-\!n_i}{\sqrt{2}\sigma}\right)\!\right)\!\left(\!1\!+\!\mbox{erf}\!\left(\frac{0.5}{\sqrt{2}\sigma}\!\right)\!\right)\!\!\bigg\}\!\left(\!1\!-\!\frac{B(n_i+\alpha_1,\alpha_2)\!+\!B(\alpha_1,n_i+\alpha_2)}{B(\alpha_1,\alpha_2)}\right).
\end{align*}
\normalsize

When $K\ge2$, DR-HA in cell $\C_i$ in the sanitized frequency distribution via the Gaussian mechanisms in Scenario 1 is 
\begin{align*}
\rho^{\text{l}}_i&=\sum_{k=1}^{K}\left(\Pr(n_{i,k}\!=\!n_i)\left(\prod_{t\neq k}\Pr(n_{i,t}\!+\!e_{i,t}\!<\!0.5|n_{i,k}\!=\!n_i)\right)\Pr(n_{i,k}\!+\!e_{i,k}\!\geqslant\!0.5|n_{i,k}\!=\!n_i)\!\right)\\
&=\sum_{k=1}^{K}\left(p_{i,k}^{n_i}\left(\prod_{t\neq k}\Pr(e_{i,t}<0.5)\right)P(e_{i,k}\geqslant0.5-n_i)\right)\\
&=\frac{1}{2^K}\left(\sum_{k=1}^{K}p_{i,k}^{n_i}\right)\left(1+\mbox{erf}\left(\frac{1}{2\sqrt{2}\sigma}\right)\right)^{K-1}\left(1+\mbox{erf}\left(\frac{n_i-0.5}{\sqrt{2}\sigma}\right)\right).
\end{align*}
The DR-HA in Scenario 8 is
\small 
\begin{align*}
\rho^{\text{l}}_i&<\mathbbm{1}(n_i\!\ge\!2)\!\left(1-\sum_{k=1}^K p_{i,k}^{n_i}\right)\bigg\{\!\Pr(e_{i,1}\geq 1.5-n_i)\Pr(e_{i,2}\geq-0.5)\prod_{t=3}^K\Pr(e_{i,t}<0.5)+\\
&\Pr(e_{i,1}< 1.5-n_i)\Pr(e_{i,2}>-0.5)\prod_{t=3}^K\Pr(e_{i,t}<0.5)\!\bigg\}\\
&=\!\mathbbm{1}(n_i\!\ge\!2)\!\left(\!1\!-\!\sum_{k=1}^K p_{i,k}^{n_i}\!\right)\!\frac{1}{2^K}\bigg\{\!\left(1\!-\!\mbox{erf}\left(\frac{1.5\!-\!n_i}{\sqrt{2}\sigma}\right)\right)\!\left(1\!+\!\mbox{erf}\left(\frac{-0.5}{\sqrt{2}\sigma}\right)\!\right)\!\left(1\!+\!\mbox{erf}\left(\frac{0.5}{\sqrt{2}\sigma}\right)\right)^{K-2}\!\!\!\!\!+\!\left(\!1\!+\!\mbox{erf}\left(\frac{1.5\!-\!n_i}{\sqrt{2}\sigma}\right)\!\right)\!\left(1\!+\!\mbox{erf}\left(\frac{0.5}{\sqrt{2}\sigma}\right)\!\right)^{K-1}\!\bigg\}\\
&=\!\mathbbm{1}(n_i\!\ge\!2)\!\left(\!1\!-\!\sum_{k=1}^K p_{i,k}^{n_i}\!\right)\!\frac{1}{2^K}\left(1\!+\!\mbox{erf}\left(\frac{0.5}{\sqrt{2}\sigma}\right)\right)^{K-2}\bigg\{\!\left(1\!-\!\mbox{erf}\left(\frac{1.5\!-\!n_i}{\sqrt{2}\sigma}\right)\right)\!\left(1\!+\!\mbox{erf}\left(\frac{-0.5}{\sqrt{2}\sigma}\right)\!\right)\!+\!\left(1\!+\!\mbox{erf}\left(\frac{1.5\!-\!n_i}{\sqrt{2}\sigma}\right)\right)\!\left(1\!+\!\mbox{erf}\left(\frac{0.5}{\sqrt{2}\sigma}\right)\!\right)\!\bigg\}.
\end{align*}
The expected DR-HA in cell $\C_i$ is
\begin{align*}
\rho^{\text{e}}_i&<\frac{1}{2^K}\left(\sum_{k=1}^{K}p_{i,k}^{n_i}\right)\left(1+\mbox{erf}\left(\frac{1}{2\sqrt{2}\sigma}\right)\right)^{K-1}\left(1+\mbox{erf}\left(\frac{n_i-0.5}{\sqrt{2}\sigma}\right)\right)+\\
&\mathbbm{1}(n_i\!\ge\!2)\!\left(\!1\!-\!\sum_{k=1}^K p_{i,k}^{n_i}\!\right)\!\frac{1}{2^K}\left(1\!+\!\mbox{erf}\left(\frac{0.5}{\sqrt{2}\sigma}\right)\right)^{K-2}\bigg\{\!\left(1\!-\!\mbox{erf}\left(\frac{1.5\!-\!n_i}{\sqrt{2}\sigma}\right)\right)\!\left(1\!+\!\mbox{erf}\left(\frac{-0.5}{\sqrt{2}\sigma}\right)\!\right)\\
&+\!\left(1\!+\!\mbox{erf}\left(\frac{1.5\!-\!n_i}{\sqrt{2}\sigma}\right)\right)\!\left(1\!+\!\mbox{erf}\left(\frac{0.5}{\sqrt{2}\sigma}\right)\!\right)\!\bigg\}.
\end{align*}
The shrinkage DR-HA for cell $\C_i$, assuming $\p_i\sim\mbox{Dirichlet}(\alpha_1,\dots,\alpha_K)$, is
\begin{align*}
\rho_i^{\text{s}}&<\left[
\sum_{k=1}^{K}\left(\int\!p_{i,k}^{n_i}\prod_{k=1}^{K}p_{i,k}^{\alpha_k-1}\frac{\Gamma(\sum_k\alpha_k+n_i)}{\Gamma(\alpha_1)\cdots\Gamma(\alpha_k+n_i)\cdots\Gamma(\alpha_K)}d\mathbf{p}\right)\!\frac{\Gamma(\sum_k\alpha_k)\Gamma(\alpha_k+n_i)}{\Gamma(\sum_k\alpha_k+n_i)\Gamma(\alpha_k)}\right]\!\times\\
&\frac{1}{2^{K}}\left(1\!+\!\mbox{erf}\left(\frac{1}{2\sqrt{2}\sigma}\right)\right)^{K-1}\left(1\!+\!\mbox{erf}\left(\frac{n_i\!-\!0.5}{\sqrt{2}\sigma}\right)\right)+\\
&\mathbbm{1}(n_i\!\ge\!2)\!\left(\!1\!-\!\left[
\sum_{k=1}^{K}\left(\int\!p_{i,k}^{n_i}\prod_{k=1}^{K}p_{i,k}^{\alpha_k-1}\frac{\Gamma(\sum_k\alpha_k\!+\!n_i)}{\Gamma(\alpha_1)\cdots\Gamma(\alpha_k\!+\!n_i)\cdots\Gamma(\alpha_K)}d\mathbf{p}\right)\!\frac{\Gamma(\sum_k\alpha_k)\Gamma(\alpha_k\!+\!n_i)}{\Gamma(\sum_k\alpha_k\!+\!n_i)\Gamma(\alpha_k)}\right]\!\right)\!\\
&\frac{1}{2^K}\left(1\!+\!\mbox{erf}\left(\frac{0.5}{\sqrt{2}\sigma}\right)\right)^{K-2}\bigg\{\!\left(1\!-\!\mbox{erf}\left(\frac{1.5\!-\!n_i}{\sqrt{2}\sigma}\right)\right)\!\left(1\!+\!\mbox{erf}\left(\frac{-0.5}{\sqrt{2}\sigma}\right)\!\right)\!+\!\left(1\!+\!\mbox{erf}\left(\frac{1.5\!-\!n_i}{\sqrt{2}\sigma}\right)\right)\!\left(1\!+\!\mbox{erf}\left(\frac{0.5}{\sqrt{2}\sigma}\right)\!\right)\!\bigg\}\\
&=\frac{1}{2^{K}}\!\left(
\sum_{k=1}^{K}\frac{\Gamma(\sum_k\alpha_k)\Gamma(\alpha_k+n_i)}{\Gamma(\sum_k\alpha_k+n_i)\Gamma(\alpha_k)}\right)\!\left(1\!+\!\mbox{erf}\left(\frac{1}{2\sqrt{2}\sigma}\right)\right)^{\!K-1\!}\!\!\left(1\!+\!\mbox{erf}\left(\frac{n_i-0.5}{\sqrt{2}\sigma}\right)\right)+\\
&\mathbbm{1}(n_i\!\ge\!2)\!\left(\!1\!-\!
\sum_{k=1}^{K}\frac{\Gamma(\sum_k\alpha_k)\Gamma(\alpha_k+n_i)}{\Gamma(\sum_k\alpha_k+n_i)\Gamma(\alpha_k)}\!\right)\frac{1}{2^K}\left(1\!+\!\mbox{erf}\left(\frac{0.5}{\sqrt{2}\sigma}\right)\right)^{K-2}\\
&\bigg\{\!\left(1\!-\!\mbox{erf}\left(\frac{1.5\!-\!n_i}{\sqrt{2}\sigma}\right)\right)\!\left(1\!+\!\mbox{erf}\left(\frac{-0.5}{\sqrt{2}\sigma}\right)\!\right)\!+\!\left(1\!+\!\mbox{erf}\left(\frac{1.5\!-\!n_i}{\sqrt{2}\sigma}\right)\right)\!\left(1\!+\!\mbox{erf}\left(\frac{0.5}{\sqrt{2}\sigma}\right)\!\right)\!\bigg\}.
\end{align*}
\end{proof}

\vspace{-24pt}
\section{\large special case of Theorems 1 and 3} \label{app:special}\vspace{-6pt}
\begin{cor}\label{cor:rho.lap}
When $K\!=\!2$,  the average local DR-HA in Theorem 1 becomes
\begin{align}
\bar{\rho}^{\text{l}}\!<&\big(1\!-\!\frac{1}{2}e^{-0.5\epsilon}\big)\frac{1}{N}\!\sum_{i=1}^N\!\bigg\{\! 
\mathbbm{1}(|\mathcal{Y}_i|\!=\!1)\big(1\!-\!\frac{1}{2}e^{(0.5-n_i)\epsilon}\big)+\mathbbm{1(|\mathcal{Y}_i|\!>\!1)}\left(e^{-0.5\epsilon}\!+\!e^{(1.5-n_i)\epsilon}\!-\!e^{(1-n_i)\epsilon}\!\right)\!\bigg\};\label{eqn:average.l}
\end{align}
the plug-in estimate of the average expected DR-HA is \vspace{-3pt}
\begin{align}
\bar{\hat{\rho}}^{\text{e}}
&\!<\!\left(1\!-\!\frac{1}{2}e^{-0.5\epsilon}\!\right)\!\frac{1}{N}\!\sum_{i=1}^N\!\left(\hat{p_i}^{n_i}\!\!+\!(1\!-\!\hat{p_i})^{n_i}\right)\!\left(1\!-\!\frac{1}{2}e^{(0.5-n_i)\epsilon}\right)\notag\\
&+\textstyle(2N)^{-1}\!\sum_{i=1}^N\!\big\{\! \mathbbm{1}(n_i\!\ge\!2)\!\left(1-\hat{p_i}^{n_i}\!-\!(1\!-\!\hat{p_i})^{n_i}\right)(e^{-0.5\epsilon}\!+\!e^{(1.5-n_i)\epsilon}\!-\!e^{(1-n_i)\epsilon})\!\big\};\label{eqn:average.e}\vspace{-3pt}
\end{align}
where $\hat{p}_i=n_{i1}/n_i$.  Assume $p_i\sim\mbox{beta}(\alpha_1,\alpha_2)$ for $i=1,\ldots,N$. The average shrinkage DR-HA  is
\begin{align}
\!\!\!\!\bar\rho^{\text{s}}&
<\!\frac{(1\!-\!\frac{1}{2}e^{-0.5\epsilon})}{B(\alpha_1,\alpha_2)N}\!
\sum_{i=1}^N\left(\!1\!-\!\frac{1}{2}e^{(0.5-n_i)\epsilon}\!\right)\!B_{1i}+\textstyle(2N)^{-1}\sum_{i=1}^{N}\big\{\mathbbm{1}(n_i\!\ge\!2)\big(1\!-\!\frac{B_{1i}}{B(\alpha_1,\alpha_2)}\big)(e^{-0.5\epsilon}\!+\!e^{(1.5-n_i)\epsilon}\!-\!e^{(1-n_i)\epsilon})\big\};\label{eqn:average.s}
\end{align}
where $B_{1i}=B(n_i+\alpha_1,\alpha_2)+B(\alpha_1,n_i+\alpha_2)$ and  $B( )$ is the beta function. The plug-in estimate of marginal DR-HA  is
\begin{align}
\bar{\hat{\rho}}^{\text{m}}
&\!<\!\left(1\!-\!\frac{1}{2}e^{-0.5\epsilon}\!\right)\!\frac{1}{N}\!\sum_{i=1}^N\!\big\{\!f(n_i;\bs\beta)\!\left(\hat{p_i}^{n_i}\!\!+\!(1\!-\!\hat{p_i})^{n_i}\right)\big(1\!-\!\frac{1}{2}e^{(0.5-n_i)\epsilon}\big)\!\big\}\notag\\
& +(2N)^{-1}\!\sum_{i=1}^N\!\big\{\! \mathbbm{1}(n_i\!\ge\!2)\!f(n_i;\bs\beta)\left(1-\hat{p_i}^{n_i}\!-\!(1\!-\!\hat{p_i})^{n_i}\right)(e^{-0.5\epsilon}\!+\!e^{(1.5-n_i)\epsilon}\!-\!e^{(1-n_i)\epsilon})\!\big\};\label{eqn:average.m}
\end{align}
and the marginal shrinkage DR-HA  is
\begin{align}
&\rho^{\text{ms}}\!<\!
\frac{(1\!-\!\frac{1}{2}e^{-0.5\epsilon})}{B(\alpha_1,\alpha_2)}\!\!\sum_{n_i=1}^{\infty}\!f(n_i;\bs\beta)\!\left(\!1\!-\!\frac{1}{2}e^{(0.5-n_i)\epsilon}\!\right)\!B_{1i}\!+\!\!\sum_{n_i=2}^{\infty}\!\frac{f(n_i;\bs\beta)}{2}\!\!\left(\!1\!-\!\frac{B_{1i}}{B(\alpha_1,\alpha_2)}\!\right)\!\! (e^{-0.5\epsilon}\!+\!e^{(1.5\!-\!n_i)\epsilon}\!-\!e^{(1\!-\!n_i)\epsilon}).\label{eqn:average.ms}
\end{align}
\end{cor}
\normalsize

\begin{cor}\label{cor:rho.gau}
In the same setting as Theorem 3, when $K=2$, the average local DR-HA  of the sanitized FD is \vspace{-4pt}
\begin{align}
\bar{\rho}^{\text{l}}<&\frac{1}{2^KN}\!\sum_{i=1}^N\!\bigg\{\! 
\mathbbm{1}(|\mathcal{Y}_i|\!=\!1)\!\left(\!1\!+\!\mbox{erf}\!\left(\!\frac{0.5}{\sqrt{2}\sigma}\!\right)\!\!\right)\!\!\left(\!1\!+\!\mbox{erf}\left(\!\frac{n_i\!-\!0.5}{\sqrt{2}\sigma}\!\right)\!\!\right)+\mathbbm{1}(|\mathcal{Y}_i|\!>\!1)E_i\big\};\label{eqn:averagel.gau.MC}
\end{align}
the plug-in estimate of the average expected DR-HA in the sanitized FD via the Gaussian mechanism is
\begin{align}
\bar{\hat{\rho}}^{\text{e}}\!<&\frac{1\!+\!\mbox{erf}\left(\frac{1}{2\sqrt{2}\sigma}\right)}{4N}\!\sum_{i=1}^N\!\left(\hat{p}_i^{n_i}\!+\!(1\!-\!\hat{p}_i)^{n_i}\right)\!\!\left(1\!+\!\mbox{erf}\left(\frac{n_i-0.5}{\sqrt{2}\sigma}\right)\right)+\!\frac{1}{4N}\!\sum_{i=1}^N\bigg\{
\!\mathbbm{1}(n_i\!\ge\!2)\!\left(1-\hat{p}_i^{n_i}\!-\!(1\!-\!\hat{p}_i)^{n_i}\right)E_i\bigg\};\label{eqn:averagee.gau.MC}
\end{align}
Assume $p_i\!\sim\!\mbox{beta}(\alpha_1,\alpha_2)$, the average shrinkage DR-HA is
\begin{align}
\bar\rho^{\text{s}}<&\frac{\!1\!+\!\mbox{erf}\left(\!\frac{1}{2\sqrt{2}\sigma}\!\right)}{4N\!\cdot\! B(\alpha_1,\alpha_2)}\!\sum_{i=1}^N\bigg\{\left(\!1\!+\!\mbox{erf}\left(\!\frac{n_i-0.5}{\sqrt{2}\sigma}\!\right)\!\right)B_{1i}\bigg\}+\frac{1}{4N}\!\sum_{i=1}^{N}\bigg\{\mathbbm{1}(n_i\!\ge\!2)\!\left(1-\frac{B_{1i}}{B(\alpha_1,\alpha_2)}\right)E_i\bigg\}\label{eqn:averages.gau.MC} ,
\end{align}
where $B_{1i}=B(n_i+\alpha_1,\alpha_2)\!+\!B(\alpha_1,n_i+\alpha_2)$. The plug-in estimate of the average marginal DR-HA is
\begin{align}
&\bar{\hat{\rho}}^{\text{m}}\!<\!\frac{1\!+\!\mbox{erf}\left(\frac{1}{2\sqrt{2}\sigma}\right)}{4N}\!\sum_{i=1}^N\!\!\bigg\{\!\!\left(\hat{p}_i^{n_i}\!+\!(1\!-\!\hat{p}_i)^{n_i}\right)\!\left(\!1\!+\!\mbox{erf} \left(\!\frac{n_i\!-\!0.5}{\sqrt{2}\sigma}\right)\!\!\right)f(n_i;\bs{\beta})\!\bigg\}\!+\!\frac{1}{4N}\!\sum_{i=1}^N\!\bigg\{\!\mathbbm{1}\!(n_i\!\ge\!2)\!\left(1\!-\!\hat{p}_i^{n_i}\!-\!(1\!-\!\hat{p}_i)^{n_i}\right)E_i\!\bigg\}\!\!\label{eqn:averagem.gau.MC}
\end{align}
and the marginal shrinkage DR-HA is
\begin{align}
\rho^{\text{ms}}\!<&\frac{\!1\!+\!\mbox{erf}\!\left(\!\frac{1}{2\sqrt{2}\sigma}\!\right)\!}{4B(\alpha_1,\alpha_2)}\!\sum_{n_i=1}^{\infty}\!\!\bigg\{\!f(n_i;\bs{\beta})\!\left(\!1\!+\!\mbox{erf}\!\left(\frac{n_i\!-\!0.5}{\sqrt{2}\sigma}\!\right)\!\!\right)\!B_{1i}\!\bigg\}\!+\!\frac{1}{4}\!\sum_{n_i=2}^{\infty}\bigg\{f(n_i;\bs{\beta})
\!\left(1-\frac{B_{1i}}{B(\alpha_1,\alpha_2)}\right)E_i\bigg\}.\label{eqn:averagems.gau.2}
\end{align}
\end{cor}

\section{\large Empirical DR-HA with Soft and Hard Thresholding of HR in  Heterogeneous Cells in Sanitized FD via the  Laplace Mechanism in the Adult Data}\label{app:thresholding}\vspace{-6pt}
\begin{figure}[!htb]
\centering\vspace{-6pt}
\includegraphics[width=0.4\textwidth]{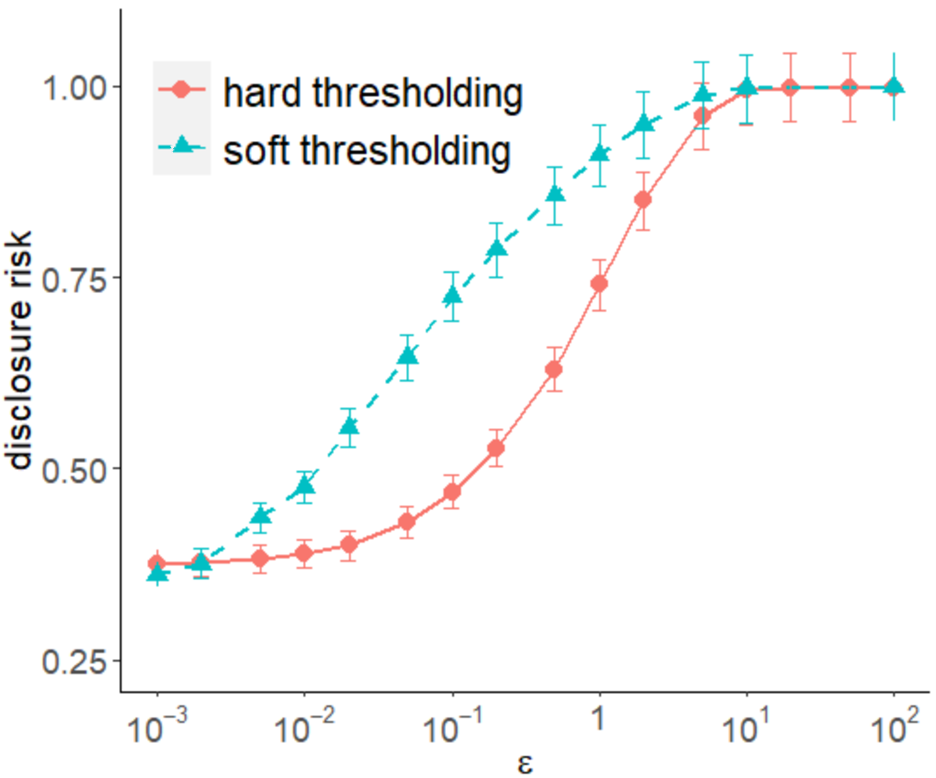}
\vspace{-6pt}
\caption{Empirical evaluation of the disclosure risk in sanitized frequency distribution in the Adult data via the Laplace mechanism}\label{fig:thresholding}
\end{figure}